\documentclass[apj,twocolumn,twocolappendix,numberedappendix]{openjournal}
\usepackage{newtxtext,newtxmath,enumitem,multirow}
\usepackage[utf8]{inputenc}
\usepackage[T1]{fontenc}
\usepackage[breaklinks,colorlinks,allcolors=blue,citecolor=blue,urlcolor=blue]{hyperref}
\usepackage{orcidlink}
\usepackage{ctable}
\pdfoutput=1 

\newcommand{\zigm}{$Z_{\rm IGM}$}
\newcommand{\etamx}{$\eta_{\rm max}$}
\newcommand{\fehtxt}{[Fe/H]}         
\newcommand{\feh}{\mathrm{[Fe/H]}}   
\newcommand{\Mv}{M_V}   

\def\yZ{y_{\rm Z}}

\def\zetaw{\zeta_{\rm w}}
\def\sfr{\dot{\mathcal{M}}_{\star}}

\def\dotMgin{\dot{M}_{\rm g,in}}

\def\Mg{M_{\rm g}}

\def\Zg{Z_{\rm g}}
\def\Zs{Z_\star}
\def\Zw{Z_{\rm w}}
\def\MZg{M_{Z,\rm g}}

\def\dotMZg{\dot{M}_{Z,\rm g}}

\def\Zigm{Z_{\rm IGM}}

\shorttitle{Metallicity of ultra-faint dwarfs}
\shortauthors{Wheeler et al.}

\begin{document}

\title{What sets the metallicity of Ultra-Faint Dwarfs?\vspace{-15mm}} 

\author{Vance Wheeler\,\orcidlink{0000-0003-4679-4435}$^{1,2, \star}$, Andrey Kravtsov\,\orcidlink{0000-0003-4307-634X}$^{2,3,4}$, Anirudh Chiti$^{2,3}$, Harley Katz$^{2,3}$, Vadim A. Semenov\,\orcidlink{0000-0002-6648-7136}$^{5}$}

\affiliation{$^{1}$Department of Physics, The University of Chicago, Chicago, IL 60637 USA}
\affiliation{$^{2}$Kavli Institute for Cosmological Physics, The University of Chicago, Chicago, IL 60637 USA}
\affiliation{$^{3}$Department of Astronomy \& Astrophysics, The University of Chicago, Chicago, IL 60637 USA}
\affiliation{$^4$Enrico Fermi Institute, The University of Chicago, Chicago, IL 60637}
\affiliation{$^{5}$Center for Astrophysics, Harvard \& Smithsonian, 60 Garden St, Cambridge, MA 02138, USA}

\thanks{$^\star$Email: \href{mailto:vanwheel@uchicago.edu}{vanwheel@uchicago.edu}}

\begin{abstract}

We use intergalactic medium (IGM) metallicity distributions from several state-of-the-art cosmological simulations of Milky Way analogs and a semi-analytic model of ultra-faint dwarf galaxy (UFD) formation to model the stellar metallicities of UFDs in MW-like environments. We study simulations with different treatments of star formation, stellar feedback, and Population III enrichment, and in all cases, we find that only a few percent of the IGM accretable by UFD progenitors is enriched to metallicities $\rm [Fe/H]\ge-4$. When the metallicity of accreted IGM in the semi-analytic galaxy formation model is set using these IGM metallicity distributions, the model underpredicts UFD metallicities and their scatter compared to the observed luminosity--metallicity relation. Our results indicate that IGM enrichment is not the dominant mechanism setting metallicities of UFD stars. Instead, UFD stellar metallicity is determined primarily by the interplay between internal enrichment and metal loss through feedback-driven outflows. We examine models with different values of the maximum outflow mass loading factor $\eta_{\rm max}$ and show that the full range of average stellar metallicities of UFDs at $M_V<-7$ can be reproduced if the maximum mass loading factor varies in the range $200\lesssim\eta_{\rm max}\lesssim 2000$. We also consider stellar metallicity distribution functions (MDFs) within individual model galaxies with different assumptions about IGM enrichment and $\eta_{\rm max}$. We find that all considered models are in reasonable agreement with observed UFD MDFs, with model differences less than the uncertainties of current metallicity measurements. 

\end{abstract}


\maketitle

\section{Introduction}
\label{sec:intro}

The advent of wide-area surveys over the past three decades resulted in an incredible expansion in the number and luminosity range of known dwarf galaxies \citep[e.g.,][]{Simon_2019,Drlica_Wagner.etal.2020}.
The discovery of the ultra-faint dwarf (hereafter UFD) galaxies \citep{Willman_etal_2005,Belokurov_etal_2007} with absolute magnitudes from $M_V\gtrsim -7.7$ \citep[or $L_V\lesssim 10^5\,L_{\odot}$,][]{Simon_2019} and extending to $M_V>0$ \citep[e.g.,][]{Smith_etal_2024} with old stellar populations \citep[e.g.,][]{Brown_etal_2012,Weisz_etal_2014,Savino_etal_2025} confirmed a general prediction based on galaxy formation simulations \citep{Ricotti_Gnedin_2005,Bovill_Ricotti_2009}, while further modeling showed that UFDs are expected to be ubiquitous in the Local Group \citep{Gnedin_Kravtsov_2006,Tollerud_etal_2008,Koposov_etal_2009,Bovill_Ricotti_2011,MK22}. Old stellar populations in UFDs provide a unique window into the early stages of evolution of these galaxies and of the Milky Way environment in which they evolve.
 
These dwarf galaxies were also recognized as powerful probes of dark matter properties on the smallest scales and useful laboratories of galaxy formation physics \citep[see, e.g.,][for reviews]{Bullock_Boylan-Kolchin_2017,Simon_Geha_2021,Sales_etal_2022}. This motivated multi-pronged observational efforts to discover and characterize the properties of the full dwarf population \citep[e.g.,][]{Belokurov_etal_2010,Koposov_etal_2015,Drlica_Wagner_etal_2015,Smith_etal_2024} and their remnants in the form of stellar streams and stellar halo stars \citep[e.g.,][]{Helmi_2020}. The upcoming Legacy Survey of Space and Time (LSST) with the Vera C. Rubin Observatory \citep{Ivezic_etal_2019} is expected to increase the number of known UFDs by an order of magnitude \citep[e.g.,][]{MK22}, which will open a new era in studying galaxy formation in this extreme regime and using the smallest galaxies to constrain properties of dark matter and galaxy formation physics. At the same time, recent and ongoing surveys and follow-up observations have already produced UFD samples that provide stringent constraints on the dark matter properties \citep[e.g.,][]{Nadler_etal_2021,Newton_etal_2021,Dalal_Kravtsov_2022} and the small-scale amplitude of the matter power spectrum  \citep{Esteban_etal_2024,Dekker_Kravtsov_2025}.
 
Properties of the currently known population of UFD galaxies have also proved to be a challenge for galaxy formation simulations. Although simulation results qualitatively agree with the old ages of stellar population, small stellar masses, and sizes of observed UFDs, they generally underpredict the average metallicities of these systems and their scatter \citep[e.g.,][]{Munshi_etal_2019,Sanati_etal_2023,Go_etal_2025}. Specifically, the luminosity--metallicity relation of observed UFDs ``flattens'' at $M_V\lesssim -7$, with average metallicities scattered by $\approx 0.5$ dex around the characteristic value of $\mathrm{[Fe/H]}\approx -2.5$ (see Figure~\ref{fig:mag-metal_mw-like} below), while the luminosity--metallicity relation in simulations extends into the UFD regime with little change in slope, relatively small scatter and average metallicities up to $\approx 1$~dex lower than observed.

Such difficulties are not surprising given the tremendous computational challenge of resolving processes on a wide range of scales required to model such tiny galaxies and their environments in the cosmological context. In addition, the modeling of star formation and stellar feedback in such simulations, calibrated using observations of more massive systems, may not extrapolate correctly to the smallest dwarf galaxies due to their shallow potential wells, low metallicities, and different dynamics and thermodynamics of their interstellar medium. 

One of the uncertainties in simulations of UFDs is the metallicity of the gas accreted from the intergalactic medium (IGM), which can be enriched by systems of different mass from mini-halos to massive galaxies. Resolving the contribution of these objects while also resolving the evolution of a UFD progenitor is usually not feasible. Likewise, star formation and stellar feedback in the UFD progenitors may proceed in a stochastic mode due to a small number of massive stars born in a typical star formation event. Thus, the modeling of both the external and internal enrichment processes in simulations of these objects is quite uncertain.  

Such difficulties motivated the development and use of semi-analytic models of UFD evolution, in which relevant processes are modeled using physically motivated or simulation-based parameterizations \citep[e.g.,][]{Salvadori_Ferrara_2009,Krumholz.Dekel.2012,Feldmann.2013,Bose_etal_2018}. The advantage of such models is their small computational expense, which allows exploration of different assumptions, evolutionary scenarios, and parameter space. 
Although these models are not a substitute for the comprehensive modeling of relevant processes afforded by full cosmological galaxy formation simulations, they can isolate processes that are particularly important in shaping specific properties of galaxies, identify the sources of discrepancies between simulations and observations, and inform simulation design in ways that can bring simulations into better agreement with observations.  

In particular, such models have been used to identify the main processes controlling the heavy element abundance in interstellar gas and stars in galaxies \citep[e.g.,][]{Finlator.Dave.2008,Peeples.Shankar.2011,Krumholz.Dekel.2012}: the inflow of metals with accreted gas from the IGM, injection of new heavy elements into the interstellar medium (ISM) by stars during their evolution, and the removal of metals in feedback-driven outflows. Enrichment by stars depends on the initial mass function (IMF) of stars and their nucleosynthetic yields. Subsequent outflow metal loss depends on feedback energy, also impacted by the IMF, and the relative metallicity of the outflowing material compared to the ISM. In addition, UFD metallicity may also be sensitive to the enrichment of the IGM or self-enrichment by Population III (hereafter Pop III) stars \citep[e.g.,][]{Wise_etal_2012,Sanati_etal_2023,Mead_etal_2025}.

The rate of outflows is often parameterized by the wind mass loading factor $\eta$ defined as the ratio of the gas mass outflow rate to the star formation rate. Galaxy formation simulations of more massive systems indicate that $\eta$ increases with decreasing stellar mass \citep[e.g.,][]{Muratov_etal_2015,Muratov.etal.2017,Angles_Alcazar_etal_2017,Mitchell_etal_2020,Pandya_etal_2021,Nelson_etal_2019}. However, the outflow properties in UFD-size galaxies are currently uncertain and it is expected that outflows in such systems are both inefficient and stochastic due to the small number of supernovae per unit of their total halo mass \citep[e.g.,][]{Penarrubia_etal_2012}. 

Currently, there is no established physical interpretation of the metallicities of UFD galaxies and their scatter in terms of the physical processes shaping the evolution of these galaxies. 
In this study, we attempt to construct such an interpretation using a combination of the IGM metallicity distributions derived from several state-of-the-art cosmological simulations of galaxy formation and a semi-analytical model of galaxy formation based on a large number of mass assembly histories of UFD progenitors residing in the Milky Way-sized halos at $z=0$.
Specifically, we explore a range of models with different assumptions in modeling the accreted IGM metallicity and different outflow mass loading factors. We compare the luminosity--metallicity relation of the model galaxies and their stellar metallicity distribution functions (MDFs) to existing measurements for observed UFDs with the goal of gauging the relative importance of external IGM enrichment and outflows in setting UFD metallicities. 

The paper is organized as follows. We outline the model, simulations and observational data used for this study in Section \ref{sec:models_and_data}. We explore the effect of the assumptions about the distribution of metallicity of gas available for accretion by UFDs and of the maximum mass loading factor of outflows on the average stellar metallicity of model UFDs in Sections \ref{sec:pre-enrichment} and  \ref{sec:outflow_modulation}, respectively. In Section \ref{sec:MDFs}, we compare the distribution of stellar metallicities within model UFDs to observed stellar metallicity distributions in several UFDs. We discuss the implications of our conclusions and their caveats in Section \ref{sec:discussion} and summarize our results and conclusions in Section \ref{sec:conclusion}.

\section{Models and Data} 
\label{sec:models_and_data}

\subsection{Model of UFD evolution}
\label{subsec:model_grumpy}

Our model for the evolution of ultra-faint galaxies is based on the halo evolution tracks extracted from the Caterpillar\footnote{\url{https://www.caterpillarproject.org}}
suite of cosmological zoom-in $N$-body simulations of Milky Way-mass ($0.5\lesssim M/10^{12}\,M_\odot\lesssim 2$) halos \citep{Caterpillar}.  Specifically, we use the $32$ LX14 Caterpillar simulations with the dark matter particle mass of $3 \times 10^4 M_\odot$. This resolution has been shown to be sufficient to reliably model subhalos down to {\it peak} masses (largest mass during halo evolution) of $\gtrsim 10^7\, M_\odot$ \citep{Pham_etal_2023}. The objects we model in the current study form in halos of $\geq 10^8\,M_\odot$, well above this limit.  Combining these suites provides a population of several thousand halos in MW-like environments, which we use in our study.

The evolution tracks of the halos in Caterpillar simulations were extracted from simulations using the Consistent Trees code \citep{Behroozi.etal.2013} and consist of  several halo properties, such as their position, virial mass, etc., measured at a series of redshifts from the first epoch at which progenitors are identified to $z=0$. 

The mass evolution tracks of halos in the zoom-in regions of the simulations described above were used as input for the \texttt{GRUMPY} regulator-type galaxy formation model  \citep{GRUMPY} based on the models of \citet{Krumholz.Dekel.2012} and \citet{Feldmann.2013}, but with a number of modifications to extend the model into the dwarf galaxy regime. This model 
evolves key properties of gas and stars (masses, metallicities, sizes, star formation rates, etc.) of the galaxies hosted by input halos by solving a system of coupled differential equations governing the evolution of these properties. Unless otherwise noted, we use the fiducial model parameter values of \citet{MK22}. 

The two aspects of the \texttt{GRUMPY} model particularly pertinent to this study are the accretion of intergalactic medium and the rate of mass and metal loss due to the feedback-driven outflows. The accretion of gas is modeled as 
$\dot{M}_{\rm g,in} = \epsilon_{\rm in}\,\Omega_{\rm b}/\Omega_{\rm m}\, \dot{M_{\rm h}}$, 
where $\Omega_{\rm b}$ and $\Omega_{\rm m}$ are the mean baryon and total matter densities in units of the critical density, and $\dot{M_{\rm h}}$ is the total mass accretion rate computed using Caterpillar halo mass evolution tracks. The factor $\epsilon_{\rm in}$ parameterizes the fraction of the universal baryon mass fraction, $\Omega_{\rm b}/\Omega_{\rm m}$, accreted by the object in gaseous form and is modeled using simulation-calibrated approximation of \citet{Okamoto_etal_2008}, as described in Section 2.2.2 of \citet{GRUMPY}.  

The heavy element (metal) mass, $\MZg$ within the ISM of model galaxies evolves as: 
\begin{equation}
\dotMZg = \Zigm\dotMgin +\left[\yZ - \left(1-\mathcal{R}+\zetaw\eta\right)\,\Zg\right]\sfr,\label{eq:mzgevo}
\end{equation}
where $\Zigm$ is the characteristic metallicity of heavy elements accreted by the galaxy from the IGM, $\yZ$ is the yield of heavy elements per unit star formation rate  
produced by young massive stars and dispersed by supernovae and AGB stars, $\mathcal{R}$ is the fraction of mass lost by stars due to winds and supernovae, assumed to be instantaneous in the model, $\Zg=\MZg/\Mg$ is the mass fraction of heavy elements in the ISM, $\eta$ is the wind mass loading factor (defined as the ratio of the gas mass outflow rate to the star formation rate) and $\zetaw$ is the wind metallicity enhancement factor $\zetaw=\Zw/\Zg$ where $\Zw$ is the metallicity of the outflow. Finally, $\dot{\mathcal{M}}_\star$ is the instantaneous star formation rate of the model galaxy. 

In the models studied here, we vary prescriptions for $\Zigm$ and $\eta$. Some models assume a fixed $\Zigm$ value, while others model $\Zigm$ variation with redshift by drawing random values from a $\Zigm$ distribution that we measure in galaxy formation simulations described below. 

In the context of the \texttt{GRUMPY} model, the outflow mass rate is set by the wind mass loading factor $\eta$, which describes how efficiently gas is removed from the galaxy by outflows relative to instantaneous star formation rate:
\begin{equation}
    \dot{M}_{\rm g,out} = \eta \dot{\mathcal{M}}_\star
\end{equation}

In cosmological simulations such as FIRE-1 \citep{Muratov_etal_2015}, FIRE-2 \citep{Pandya_etal_2021}, EAGLE \citep{Mitchell_etal_2020}, and Illustris TNG50 \citep{Nelson_etal_2019}, a clear trend of increasing average $\eta$ with decreasing stellar mass is present. However, these analyses do not probe galaxies with stellar mass as low as UFDs $(M_\star \lesssim 10^{5.5} M_\odot)$, so the extent to which this relation can be extrapolated to galaxies of stellar mass $M_\star<10^6\, M_\odot$ is unclear. Furthermore, the mass loading factor of galaxies at high redshifts where UFDs are expected to form most of their stars is also largely unknown.  

The parametrization for mass loading factor adopted in \texttt{GRUMPY} is given by:
\begin{equation} \label{eq:eta}
    \eta = {\rm min}\left(\eta_{\rm max},\,\eta_{\rm norm} M_{\star,10}^{\eta_p} - \eta_c\right)
\end{equation}
The fiducial values of \citet{MK22} adopted in this work are $\eta_{\rm norm} = 1.8$, $\eta_{p} = -0.45$, $\eta_{c} = 4.5$, $M_{\star,10} \equiv M_\star / 10^{10}M_\odot$, and $\eta_{\rm max}=2000$. These values were chosen to be close to those in the FIRE-1 and FIRE-2 simulations for galaxies of $10^6\lesssim M_\star/M_\odot\lesssim 10^9$ \citep{Muratov_etal_2015,Pandya_etal_2021}, while at smaller and larger masses the power law relation $\eta\propto M_{\star,10}^{\eta_p}$ is modified to reproduce the observed stellar mass to halo mass relation (SHMR) and stellar mass--metallicity relations, as detailed in \citet{MK22}. The subtraction of $\eta_{\rm c}$ results in the rapid decrease of $\eta$ for $M_\star\gtrsim 10^{10}\, M_\odot$, while the maximum value of $\eta_{\rm max}=2000$ was chosen to account for the expectation in the UFD regime the decreasing stellar mass-to-halo mass ratio results in less efficient driving in outflows and to prevent unlimited increase of $\eta$ at the very low stellar masses. 

The values of other parameters are set to the fiducial values adopted in \citet{MK22}: $y_Z=0.06$ for the \citet{Chabrier.2003} IMF \citep[e.g.,][]{Vincenzo.etal.2016}, $\mathcal{R}=0.34$, $\zetaw=1$. The expected value and scaling of $\zetaw$ with system mass are currently not known. \citet{MK22} showed that a good match to the observed luminosity--metallicity relation of bright dwarf galaxies can be obtained with $\zetaw=1$, as long as a mass-dependent scaling of $\eta$ is adopted. Results of the FIRE-1 and FIRE-2 galaxy formation simulations are also consistent with $\zetaw\approx 1$ independent of galaxy mass \citep[][]{Muratov.etal.2017,Pandya_etal_2021}, albeit with a sizable scatter. 

Magnitudes of \texttt{GRUMPY} model galaxies are calculated using the flexible stellar population synthesis model (FSPS) \citep{Conroy_etal_2009_fsps,Conroy_Gunn_2010_fsps}. Post processing with the chemical evolution model Chempy \citep{Chempy} is used to track the evolution of  individual elements within single stellar populations (SSPs) in each galaxy. This model includes time-dependent enrichment and yields for individual elements. The output of this model is used to generate distributions of individual stellar population metallicities with which we create MDFs in Section \ref{sec:MDFs}.

\subsection{Galaxy formation simulations}

To explore the effect of the abundance of heavy elements in the IGM on the metallicity distribution of stars in dwarf galaxies, we examine several models for the metallicity of accretable gas in the context of the dwarf galaxy evolution described above. To this end, we construct metallicity distributions of the IGM gas that would be accretable by UFD progenitors at high $z$ by sampling the IGM from several high-resolution cosmological hydrodynamical simulations of MW-like galaxies and their environments. These models vary in their prescriptions for star formation, feedback, chemical enrichment and evolution, and inclusion or omission of modeling of Pop III feedback and enrichment.

We consider the properties of these simulated MW-like environments in the redshift range $5 < z < 10$, corresponding to the redshift range of the publicly available FIRE-2 high redshift suite. We found that the metallicity of the IGM at $z\approx 8-10$ has the most effect on the stellar metallicities of model UFDs within this range (see Section \ref{sec:pre-enrichment}). We use the $z=10$ metallicity distribution to model IGM metallicity in the \texttt{GRUMPY} model at $z>10$ -- an overestimate of the metallicity of accreted gas, but a conservative choice that strengthens our conclusions below. Similarly, we extend the $z=5$ metallicity distribution to $z=0$ in the \texttt{GRUMPY} model, where the choice of $\Zigm$ has little impact on model UFD metallicities because UFDs are no longer star forming.

\subsubsection{FIRE-2 simulations}
\label{subsec:model_fire}

We use simulations from the FIRE-2 public data release \citep{Wetzel_etal_2023_FIRE_pubdata}. The FIRE-2 cosmological zoom-in simulations of galaxy formation are part of the Feedback In Realistic Environments (FIRE) project, generated using the Gizmo code \citep{Hopkins_2015_GIZMO_code} and the FIRE-2 physics model \citep[][see also \S 2.1 in \citealt{Wetzel_etal_2023_FIRE_pubdata} for a succinct summary]{Hopkins_etal_18_FIRE_code}.

We consider simulation outputs in the redshift range $5 < z < 10$, the range publicly available in both the core m12 models \citep{Wetzel_etal_2016,Garrison_Kimmel_etal_2017,Garrison_Kimmel_etal_2019,Hopkins_etal_18_FIRE_code,Samuel_etal_2020} and the high redshift simulations of smaller-mass halos \citep[e.g.,][]{Ma_etal_2018,Ma_etal_2019,Ma_etal_2020}, suitable to probe the environments of UFDs prior to cosmic reionization when most of their stars are expected to form. All FIRE-2 simulations analyzed here assume an IGM gas metallicity floor of $\feh \simeq -3.9$.

\subsubsection{Semenov et al. 2025 simulation of a Milky Way progenitor}
\label{subsec:model_mw-art}

We also study a simulation of a progenitor of a Milky Way sized galaxy selected from the sample of MW progenitors from the Illustris TNG50 simulation \citep{Pillepich_etal_2018_TNG,Nelson_etal_2019_TNG_data_release} with an early forming disk resembling the chemo-kinematic structure of the MW stellar disk \citep{Semenov_etal_MW-ART}. The galaxy was re-simulated using the ART code at higher resolution and including a detailed modeling of multiphase ISM, radiative and stellar feedback, subgrid turbulence, and a model of variable star formation efficiency  \citep{Kravtsov_1999_ART,Rudd_etal_2008_ART,Gnedin_Kravtsov_2011_ART,Semenov_etal_2016,Semenov_etal_2018}. The model details are described in \citet{Semenov_etal_MW-ART}, and we refer to this simulation as S25~ART. No IGM gas metallicity floor was assumed in this model. 

\citet{Semenov_etal_2025} present a detailed analysis of the early evolution of the MW progenitor in this simulation and show that at $z\gtrsim 6-7$ the progenitor has a very bursty star formation and drives significant outflows. In this work, we study the metallicity distribution of the IGM gas accretable by UFD progenitors within the zoom-in region of this simulation at redshifts $5\leq z\leq 10$.

\subsubsection{MEGATRON simulations}
\label{subsec:model_megatron}

MEGATRON  is a high-resolution cosmological zoom-in simulation of a MW-sized galaxy progenitor selected to have a particularly early disk formation that includes a suite of physical process modeling introduced by \citet{Katz_etal_2024_MEGATRON}. The simulation was run using the RAMSES-RTZ code based on the RAMSES \citep{Teyssier_2002_RAMSES} and RAMSES-RT \citep{Rosdahl_etal_2013_RRT,Rosdahl_Teyssier_2015_RRT} codes. No metallicity floor in the IGM gas was assumed in the MEGATRON simulation. Instead, chemical enrichment is modelled self-consistently, including the contribution of Pop III stars.

We study three high-resolution MEGATRON simulations with different stellar feedback energy and stellar initial mass function (IMF) prescriptions. Full descriptions of the model parameters and differences between simulations will be described elsewhere (Katz, H. et al. in prep). It includes the Efficient SF, Bursty SF, and Variable IMF re-simulations that we analyze here. These re-simulations differ substantially in their star formation histories, burstiness of star formation, and strength of feedback and outflows. 
With the Efficient SF model as reference, the Bursty SF increases type II SNe feedback energy from $1 \times 10^{51}\ \rm erg$ to $5 \times 10^{51}\ \rm erg$, resulting in galaxies with more bursty SF histories. The Variable IMF model allows for density and metallicity dependent IMF and Pop II hypernovae with mass dependent energy.

At the current time, only the snapshots at $z \gtrsim 8.5$ are available and are used in our analyses. As noted above, for the FIRE-2 and S25~ART simulations, due to truncation of star formation due to reionization, the IGM metallicity distribution at $z\gtrsim 8$ has the greatest impact on the metallicity distribution of stars in model UFD galaxies.

\subsection{Observational data}
\label{subsec:observation}

We make use of data from several observational studies of dwarf galaxies. From measurements of Milky Way dwarf galaxies collected in the Local Volume Database\footnote{\url{https://github.com/apace7/local_volume_database}}
\citep[][release v1.0.3\footnote{\url{https://doi.org/10.5281/zenodo.14795252}}]{Pace_lvdb},
we consider the average stellar metallicities of dwarf galaxies spectroscopically determined using more than 5 stars.

We also make metallicity distribution functions from individual stellar metallicities of Bo{\"o}tes I, Hercules, Centaurus I, Leo IV, and Hydrus I, reported in 
\citet{Simon_Geha_2007_MDF,Brown_etal_2014_MDF,Koposov_2018_MDF,Jenkins_etal_2021_MDF,Longeard_etal_2022_MDF,Waller_etal_2022_MDF,Heiger_etal_2024_MDF}.
These UFDs were used because there are a relatively large number of stellar metallicity measurements (i.e., $N > 20$) for each, and they do not have significant overlap in $M_{\rm V}$, which we use to select analogous model galaxies.

\section{Results}

\subsection{Enrichment of the IGM gas and its impact on the stellar metallicity of UFDs}
\label{sec:pre-enrichment}
 
We first consider the metallicity of the IGM that would be accretable by UFDs in a MW-like environment at redshifts $5<z<10$ where they are expected to be star-forming. The goal is to gauge whether the enrichment of accretable IGM gas that is predicted by hydrodynamical simulations reproduces the mean stellar metallicity and the metallicity scatter of observed UFDs around the Milky Way.

To this end, we model the evolution of UFDs using the \texttt{GRUMPY} fiducial model and mass assembly histories of the Caterpillar halos, but assuming models with a probability distribution function (pdf) of the accreted IGM metallicity, \zigm, instead of the constant \zigm\ value of the fiducial model. We adopt the pdfs for IGM metallicity based on the gas that is accretable by UFD progenitors in several cosmological hydrodynamic simulations of the MW progenitor galaxies at $z\gtrsim 5$. 

\subsubsection{Defining the gas accretable by UFDs}

To determine the metallicity of gas accretable by UFDs across redshifts, we first need to identify gas that UFD progenitors can potentially accrete. To this end, we use Caterpillar evolutionary tracks of model UFDs, defined as galaxies with $\Mv > -7.7$ at $z=0$, following the UFD limit adopted by \citet{Simon_2019}. We also limit analysis to the dwarfs that have apparent $V$-band magnitudes of $m_V < 17.5$ if observed from the center of their parent halo at $z=0$, which approximately corresponds to the apparent magnitudes of UFDs identified in SDSS, DES, and PanSTARRS surveys \citep[see][]{Drlica_Wagner.etal.2020}. This limits the model galaxies to the luminosities of observed galaxies sufficiently bright for identification and metallicity measurements. We also restrict galaxies to a distance of $r \leq 100\ \rm kpc$ from the center of the host halo at $z=0$. This was done originally to alleviate concerns that the analyzed  Lagrangian regions in some cosmological simulations are small and also because most observed UFDs are located within 100 kpc of Milky Way.
However, we found that the size of the analyzed region has little effect on the distributions of UFD progenitor galaxy properties at the redshifts we sample. Furthermore, the closest surviving objects tend to be at smaller distances to the progenitor of the host galaxy at high $z$, which are generally more enriched. This makes the main conclusion of our study conservative. 

We use the tracks of the model galaxy sample defined in this way to identify locations of their progenitors at $5<z<10$ relative to the main progenitor of the host MW-sized galaxy and their virial masses.

\begin{figure*}
    \centering
    \includegraphics[height=0.222\textheight,trim=0 0 10 0, clip]{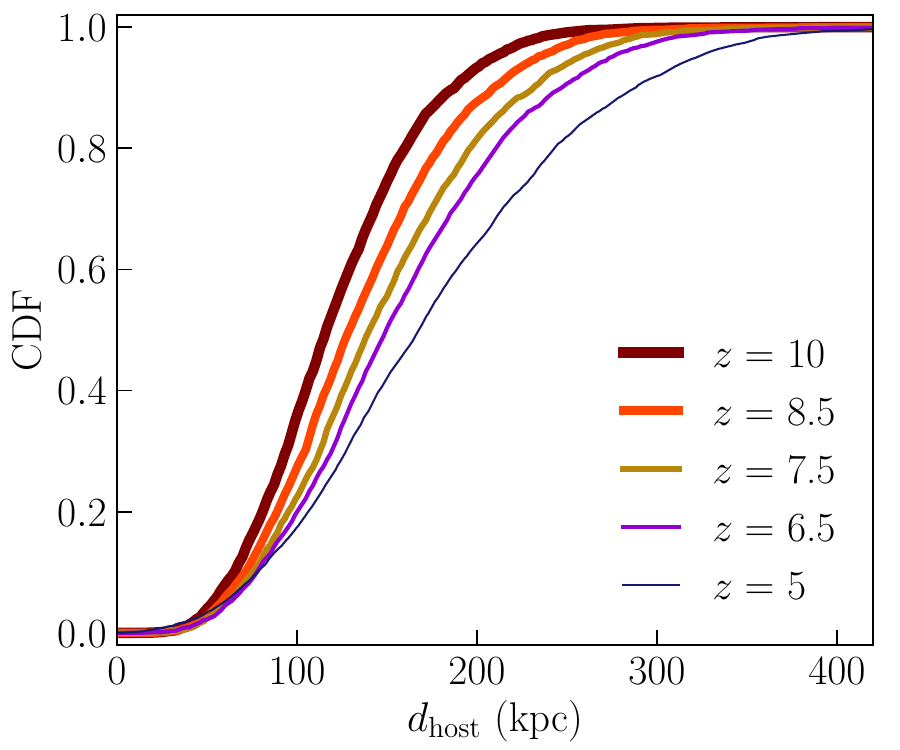}
    \includegraphics[height=0.222\textheight,trim=22 0 10 0, clip]{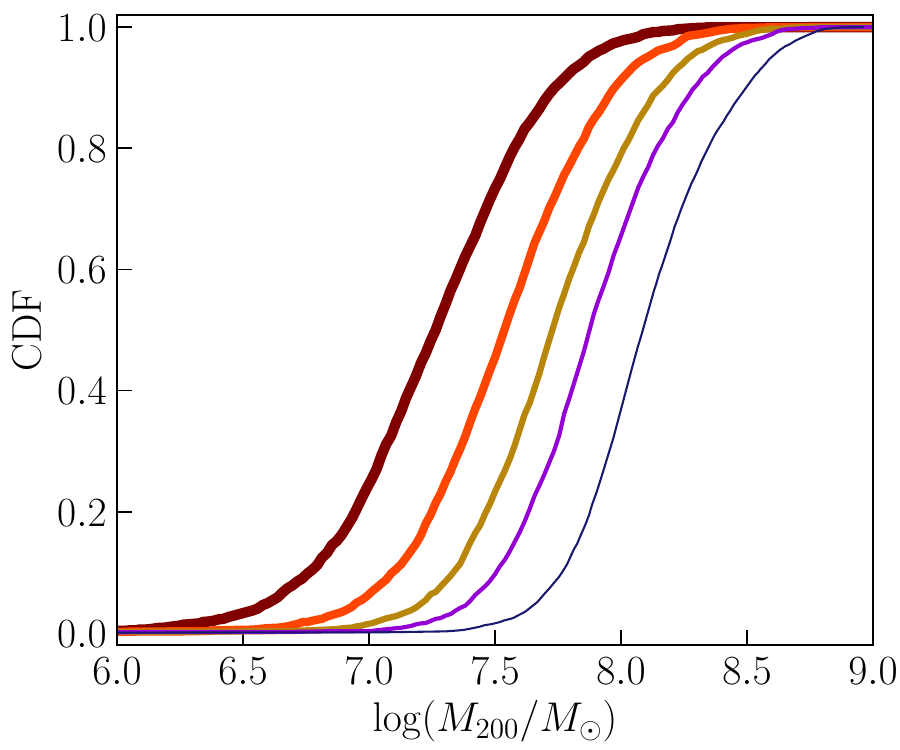}
    \includegraphics[height=0.222\textheight,trim=22 0 10 0, clip]{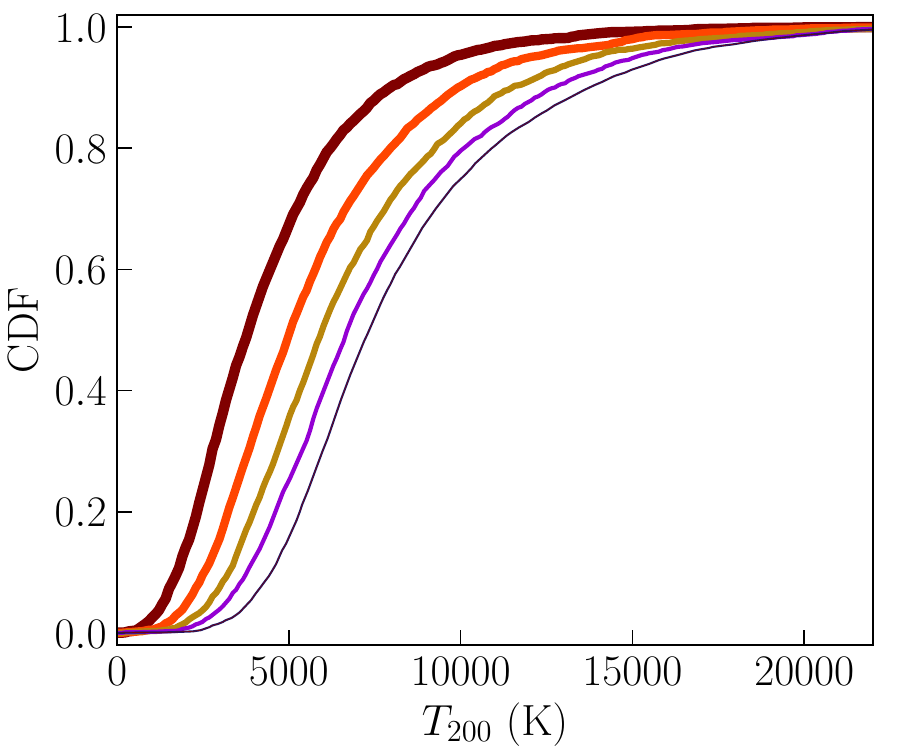}
    \caption{Cumulative distributions of the physical distance to the MW host progenitor (left), virial mass (center), and virial temperature (right) at a redshift $z$ (see legend) of the Caterpillar+\texttt{GRUMPY} UFD progenitors. The properties of mock UFD halos used to sample \zigm\ in cosmological simulation are drawn from these distributions. $T_{200}$ is derived from $M_{200}$, but is included to illustrate the temperature range of gas UFD progenitors may accrete.}
    \label{fig:ufd_progenitor_props}
\end{figure*}

Figure \ref{fig:ufd_progenitor_props} shows the cumulative distribution function of the distance to the host progenitor, UFD progenitor halo mass $M_{200}$, and the corresponding virial temperature \citep[e.g.,][]{Okamoto_etal_2008}:
\begin{equation}
    T_{200} = \frac{1}{2} \frac{\mu m_p}{k_{\rm B}} \frac{GM_{200}}{R_{200}},
\end{equation}
where the mean molecular weight is $\mu=0.59$, appropriate for ionized IGM gas. Because hot outflows may be significantly more enriched than the average IGM, the temperature of gas may have significant impact on the metallicity of gas available to UFD progenitors.

We use distances to the host progenitor, $M_{200}$, and $T_{200}$ values sampled from the distributions shown in Figure \ref{fig:ufd_progenitor_props} and gas properties in galaxy formation simulations to identify gas accretable by UFD progenitors. Specifically, we consider a gas parcel to be accretable by a UFD if 1) its density is below  $100\ \overline{\rho}_{\rm b}(z)$, thereby excluding gas within bound structures, 2) its temperature is lower than $T_{200}$, and 3) it is located within $R_{200}$ of the mock UFD progenitor. We then calculate the mass-averaged metallicity of the gas parcel, and create a metallicity distribution of the accretable gas by repeating the process for tens of thousands of mock halo locations and properties at each redshift. 

As we noted in Section~\ref{subsec:model_grumpy}, \citet{Pham_etal_2023} showed that the LX14 Caterpillar simulations  reliably model subhalos down to peak masses of $\gtrsim 10^7\, M_\odot$. Although some of the objects in  Figure \ref{fig:ufd_progenitor_props} have smaller masses at high $z$, their peak masses are $>10^8\,M_\odot$ and are well inside the regime reliably resolved in the simulations.

\subsubsection{Metallicity distribution of the IGM accretable by UFDs}
\label{subsec:accretable gas}

\begin{figure*}
    \centering
    \includegraphics[height=0.22\textheight,trim= 0 0 3 0, clip]{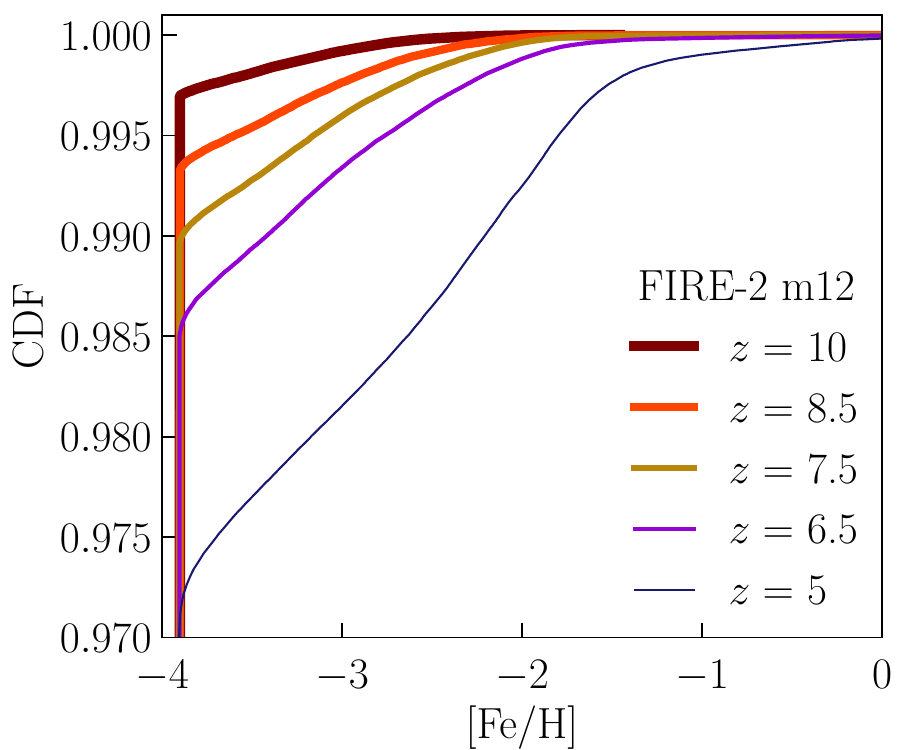}
    \includegraphics[height=0.22\textheight,trim= 29 0 3 0, clip]{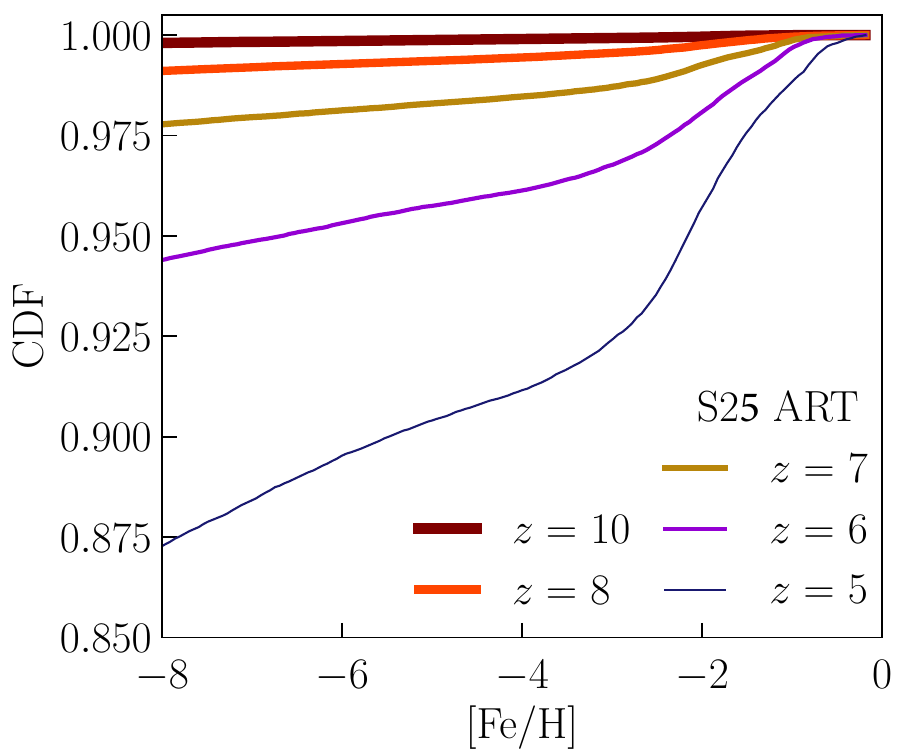}
    \includegraphics[height=0.22\textheight,trim= 29 0 3 0, clip]{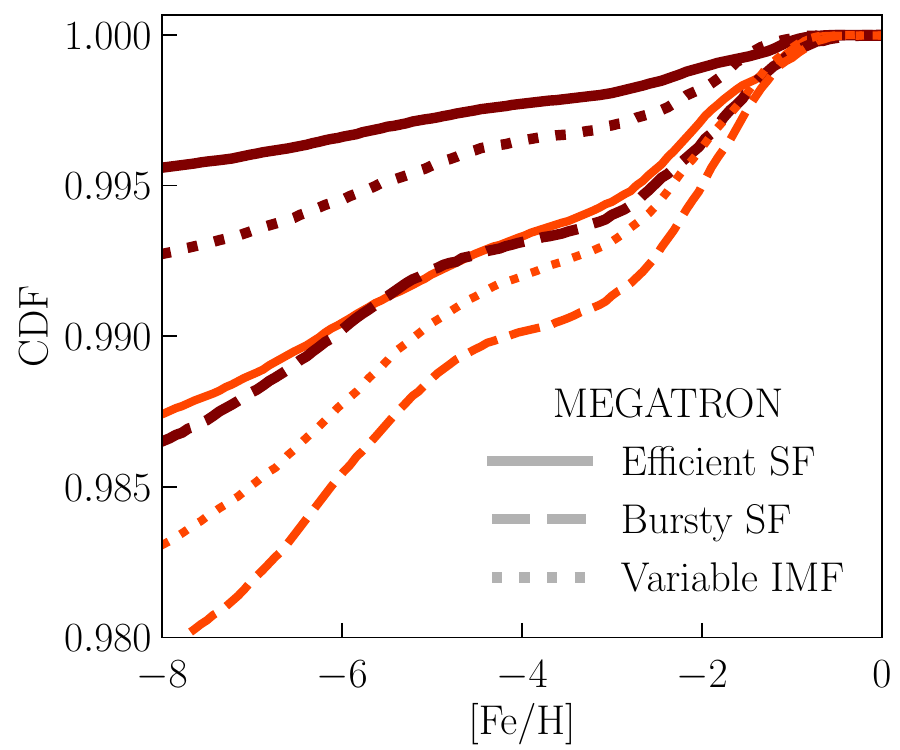}
    \caption{Cumulative distributions of the metallicity of gas available for accretion by UFD progenitor halos in the environment around simulated MW analogs at several redshifts. Average stellar metallicities in the \texttt{GRUMPY} model are taken to be defined by the mass of metals and total mass in stars relative to solar. Note that the axis limits vary from panel to panel, however in all model families, only a small fraction of enriched gas is available at the highest and most impactful redshifts.
    \textit{Left}: All 8 FIRE-2 m12 simulations, where model-to-model variation in gas availability is sufficiently low that the sample data from all models are combined to create this distribution. These models impose a metallicity floor of $\feh \simeq -3.9$.
    \textit{Center}: The S25~ART model. Unlike FIRE-2 models, no metallicity floor imposed in the model. The lower {\fehtxt} limit of this figure is selected for convenience of visualization to show the distribution in a region below $\feh = -4$, however the majority of available gas resides at significantly lower metallicities. Proportionally more gas than for other MW-like models is enriched above $\feh=-2$ as this model evolves, however at $z=10$ it is the least enriched.
    \textit{Right}: Several MEGATRON models with different feedback or star formation prescriptions are shown with different line styles. Distributions for $z=10$ and $z=8.5$ are shown with the same color scheme as FIRE-2 legend. Lower redshift data is not yet available at the time of writing. Unlike other models shown, MEGATRON models include explicit Pop III stellar prescriptions in addition to no imposed metallicity floor. In general, increasing feedback strength or changing the IMF leads to more available metals in gas near the host galaxy progenitor compared to the fiducial (Efficient SF) model. Similar to the center panel, the majority of available gas resides at significantly lower metallicity than shown.}
    \label{fig:zigm_cdf}
\end{figure*}

Figure \ref{fig:zigm_cdf} shows the cumulative distribution of the metallicity of the gas accretable by UFDs. The left panel presents distributions for all 8 FIRE-2 {\tt m12} simulations. We find that model-to-model variation of the distributions is smaller than their redshift evolution. We thus combined accretable gas from all eight simulations to construct the distribution for each redshift.  The figure shows that most of the UFD-accretable gas in the FIRE-2 simulations has rather low metallicity  with only $1\%$ of gas above $\feh \approx -4$. The range of metallicity of the accretable gas increases with decreasing redshift, but even by redshift $5$, where reionization is expected to have suppressed gas accretion thereby quenching star formation in UFDs, only a few percent of gas is enriched above $\feh \approx -4$.

The middle panel of Figure \ref{fig:zigm_cdf} shows the metallicity distribution in the S25~ART model of an early forming MW analog. The plot shows that in the S25 simulation 98\% of the gas available to UFDs at $z>7$ has $\rm [Fe/H]<-4$. At $z\approx 5-7$, however, the amount of enriched gas increases rapidly with $\approx 10\%$ of the accretable gas at $\rm [Fe/H]>-4$ at $z=5$. In addition, the range of metallicity of accretable gas in this simulation is wider than in the FIRE-2 simulations and there is a more prominent tail of gas extending to metallicities $\rm [Fe/H]>-2$ than in other simulations.

The right panel of Figure \ref{fig:zigm_cdf} shows cumulative distributions of \zigm\ at $z\geq 8.5$ in several MEGATRON re-simulations of a progenitor of MW-sized galaxy with varying feedback and star formation prescriptions, including explicit modeling of enrichment due to Pop III stars that are not included in the FIRE-2 and S25 simulations. These simulations also do not assume any minimum value of the gas metallicity.

Of the MEGATRON simulations analyzed, the gas accretable by UFDs has the lowest metallicities in the efficient SF simulations. The gas has the broadest range of metallicities in the bursty SF model, and therefore the greatest proportional availability of higher metallicity gas, reflecting the stronger outflows in this simulation. The width of the metallicity distribution in the variable IMF model is intermediate between the efficient SF and bursty SF simulations. The type II supernova energy in this simulation is the same as in the efficient SF run, but a metallicity and density-dependent IMF that becomes top-heavy in some regime, and inclusion of energy injection from hypernovae, lead to increased enrichment. The differences between these models is minor, however, and in our analyses below we consider only the bursty SF model.

All of the simulations examined here have strong feedback-driven outflows and/or an early MW progenitor formation time, as explicitly chosen in MEGATRON and S25~simulations. In addition to strong feedback ejecting more metals from the host, early forming galaxies have had more time to evolve and enrich their environments by a given redshift. Therefore the level of gas enrichment in these simulations may be considered a conservative upper limit on what can be expected for the Milky Way environments at $z>5$. Nevertheless, as we show below, this level of enrichment has a negligible effect on the metallicity distribution of stars in UFDs in the \texttt{GRUMPY} model. 

\subsubsection{Effect of the accreted IGM metallicity on the average stellar metallicity in UFDs}
\label{subsec:zigm_mag-metal}

Results in the previous section showed that the gas accretable by UFDs has generally low metallicity and a fairly narrow metallicity distribution in a diverse set of zoom-in galaxy formation simulations with efficient feedback and outflows. Here we examine whether this metallicity distribution of the accreted gas expected from simulations can reproduce the form and scatter of the observed luminosity--metallicity relation in the UFD regime.

To this end, we sample the \zigm\ distributions constructed in the previous section, interpolated between redshifts, for every point in the mass assembly history which \texttt{GRUMPY} samples $(\Delta a\ {\sim}\ 0.014\ )$ for each \texttt{GRUMPY} model galaxy. The sampled values are taken as the metallicity of the accreted gas for that period, instead of assuming a constant value as in the fiducial model calculation. Therefore, the metallicity of accreted IGM varies with redshift, and from galaxy to galaxy.

\begin{figure*}
    \centering
    \includegraphics[height=0.332\textheight,trim= 20 20 20 20, clip]{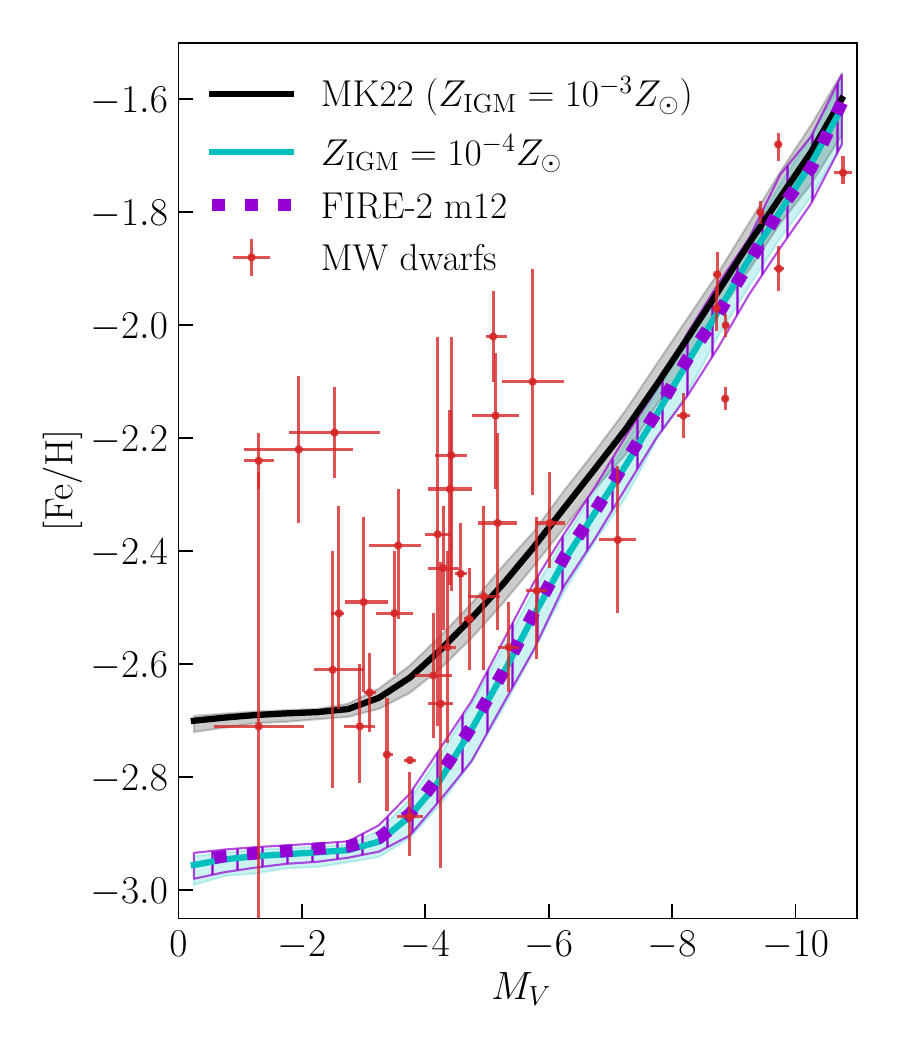}
    \includegraphics[height=0.332\textheight,trim= 82 20 20 20, clip]{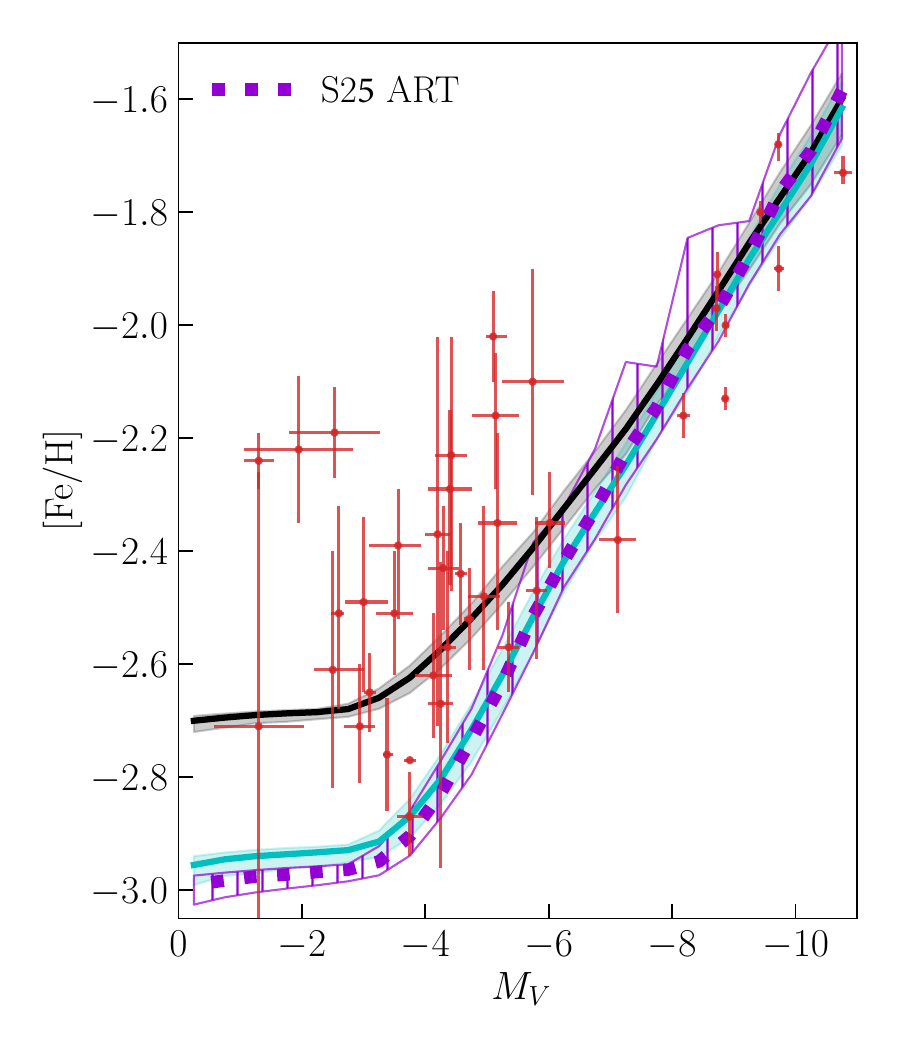}
    \includegraphics[height=0.332\textheight,trim= 82 20 20 20, clip]{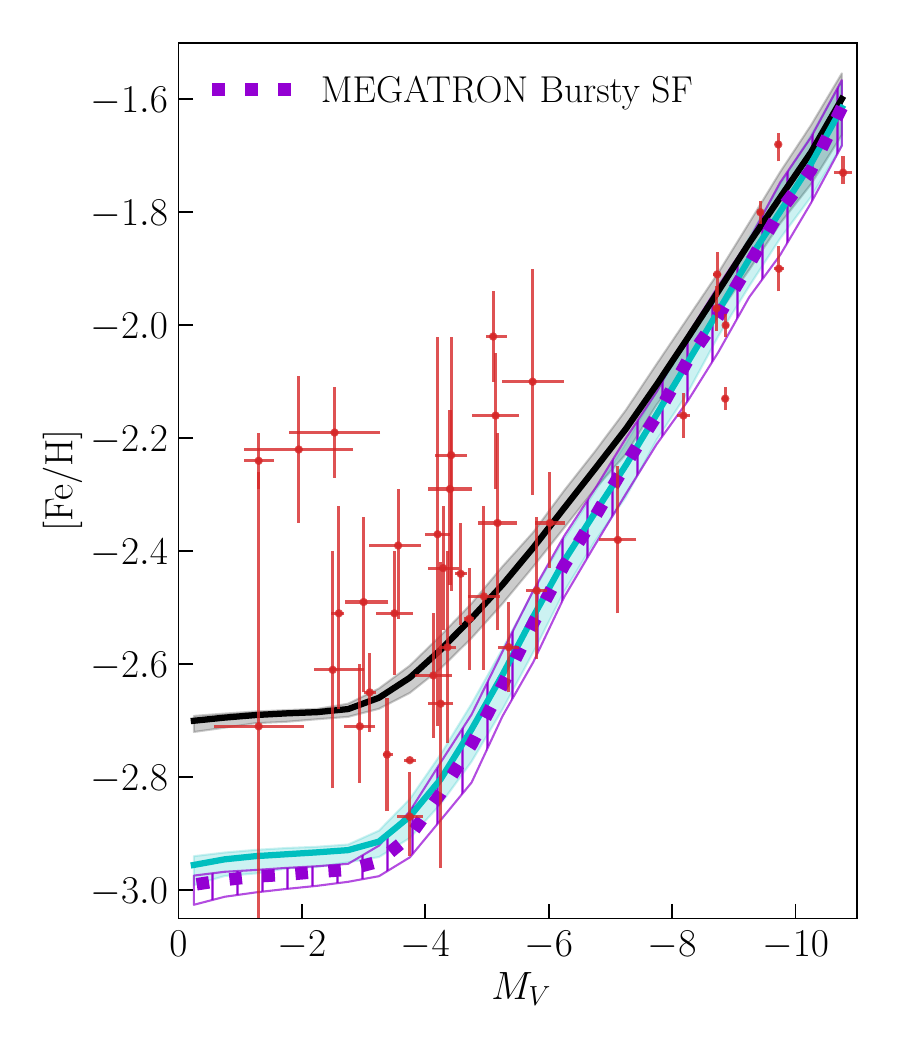}
    \caption{Average stellar metallicity vs V-band magnitude for observed and model galaxies in this study. Lines are the median, and shaded or hatched regions contain $68\%$ of model galaxies for a given $\Mv$. Red points are observed MW dwarf galaxies reported in \citet{Pace_lvdb} which have mean spectroscopic metallicities derived using $>5$ stars. All model parameters other than \zigm\ are the fiducial values of \citet{MK22}. For models informed by cosmological simulations, shown in violet, \zigm\ is drawn from the distribution categorized in Section \ref{subsec:accretable gas}. The MK22 and \zigm$=-4$ models use a constant value. All simulation-informed models, which have very little enriched gas available to UFDs, possess lower average metallicity than observed UFDs, and the intrinsic scatter in all models is significantly less than the intrinsic scatter of observed galaxies. \zigm$=-4$ is included to demonstrate the behavior of a typical metallicity floor, and all simulation-informed models very closely follow this behavior in both the average and scatter regardless of metallicity floor or prescription.}
    \label{fig:mag-metal_mw-like}
\end{figure*}

Figure \ref{fig:mag-metal_mw-like} shows the resulting average stellar metallicities as a function of galaxy $V$-band absolute magnitude. In this section and \ref{sec:outflow_modulation}, we define model metallicity by $\Zs$ interchangeably with $\feh$ using the \texttt{GRUMPY} base model as in \citet{GRUMPY,MK22}, noting that the model instantaneously recycles stellar yields. This choice does not effect our conclusions, as at early times [Fe/H] would be lower than Z because of the low contribution of SNIa, strengthening our results. The metallicity of the accreted gas has little effect for bright ($\Mv<-7$) galaxies, but affects the average metallicity of UFD galaxies substantially. In particular, models with the IGM metallicity sampled from the simulation distributions produce metallicities considerably lower than the fiducial model with constant $Z_{\rm IGM}=10^{-3}\, Z_\odot$ and lower than the typical metallicities of observed UFDs. The relations of these models can be matched well by the model with constant $Z_{\rm IGM}=10^{-4}\, Z_\odot$. In fact, the predicted ${\rm [Fe/H]}-\Mv$ relation is insensitive to the IGM metallicity or its distribution for $Z_{\rm IGM}\lesssim 10^{-4}\, Z_\odot$.

\begin{figure}
    \centering
    \includegraphics[width=1.0\linewidth,trim= 16 16 16 16, clip]{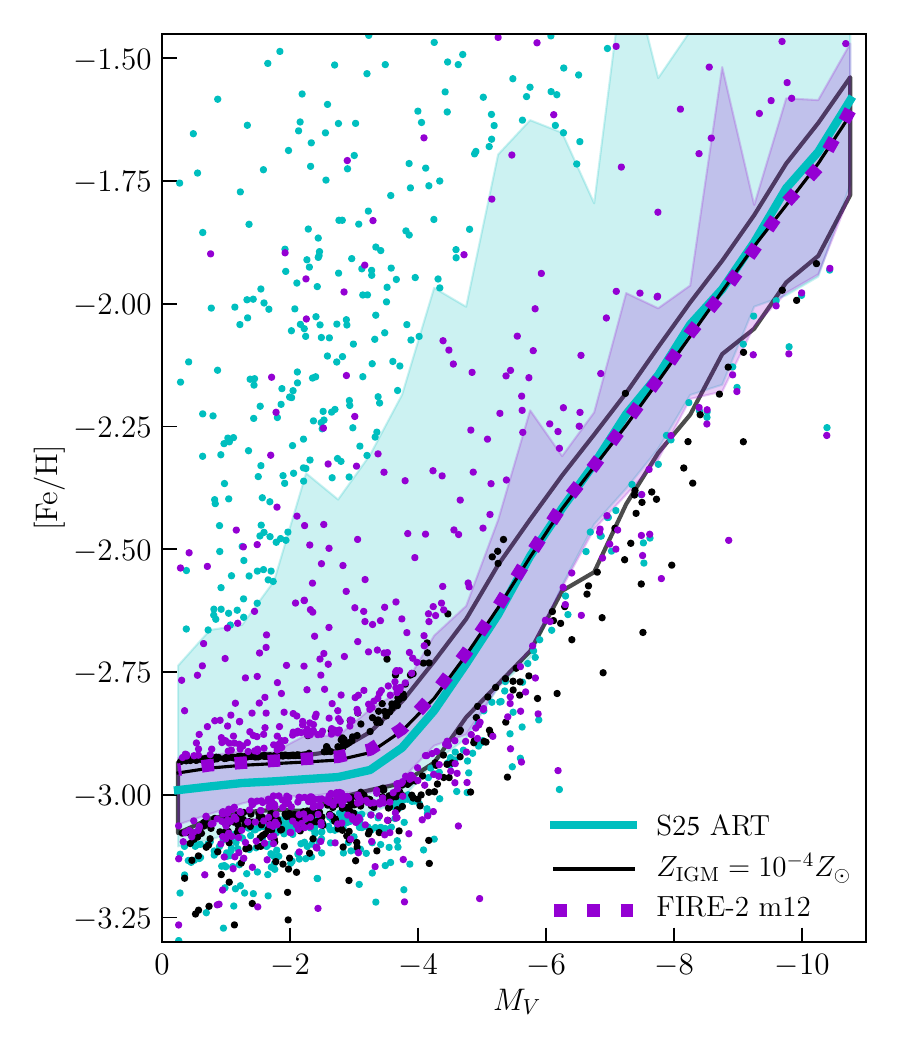}
    \caption{Metallicity vs $V$-band absolute magnitude relation of galaxies in the models that use S25 and FIRE-2 IGM metallicity distributions and in the model with constant $Z_{\rm IGM}=10^{-4}\, Z_\odot$. Lines show the medians of model galaxies in bins of absolute magnitude and shaded bands contain $95\%$ of galaxies in a particular model; galaxies outside the 95\% bands are shown as individual dots. This figure shows that the luminosity--metallicity relation is generally insensitive to the form of $Z_{\rm IGM}$ distribution for $Z_{\rm IGM}<10^{-4}\, Z_\odot$. However, the simulation-informed models produce a tail of galaxies enriched considerably more than the median for their luminosity, which is lacking in the constant $Z_{\rm IGM}=10^{-4}\, Z_\odot$ model. Likewise, the S25-informed model produces a larger fraction of galaxies in the high-metallicity tail than the FIRE-2 informed model because \zigm\ in the former is wider and extends to higher metallicities (see Figure \ref{fig:zigm_cdf}).}
    \label{fig:mag-metal-scatter}
\end{figure}

Finally, the models predict a much smaller scatter about the median ${\rm [Fe/H]}-\Mv$ relation than exhibited by observed UFD galaxies. 
Figure \ref{fig:mag-metal-scatter} shows that although the luminosity--metallicity relation for model galaxies with \zigm\ sampled from cosmological simulations is consistent with the constant $\Zigm = 10^{-4}\, Z_\odot$ models, drawing \zigm\ from these extended distributions does produce differences in the tails. Simulation-informed \zigm\ models can produce a galaxy population with a wide range in metallicities, but the number of such galaxies is small enough that it contributes to a far-reaching but weak tail of an otherwise very narrowly peaked distribution inconsistent with the observed scatter in UFD metallicities. 

The tail for S25 \zigm\ models is significantly stronger than that of FIRE-2 m12 \zigm\ models, due to the greater contribution of $\feh \gtrsim -2$ gas available to UFDs in the S25 simulation. By selectively choosing which redshifts to sample from this distribution, it was found that the metallicity of fainter galaxies near the plateau are almost entirely dictated by the metallicity distribution at redshifts above the fiducial reionization redshift $z \approx 8.5$, while the extent of brighter galaxy metallicity was similar to that of FIRE-2 m12 \zigm\ models. Therefore, it is the extremely broad S25 \zigm\ distribution at lower redshifts which contributes to the wide tail for $\Mv < -5$ galaxies shown in Figure \ref{fig:mag-metal-scatter}. The metallicity distributions in MEGATRON simulations are narrower than in S25, but produce similar metallicity scatter as described for sampling only the high redshift S25 $\Zigm$ distribution. However, MEGATRON simulations at lower redshifts are not yet available to compare over the full redshift range.

Note that the tail of galaxies with metallicities smaller than the median is generally small in all of the simulations. This is not due to the characteristic metallicity of the accreted IGM, which is far lower than the metallicities in this tail in most of the considered models. Rather, the tail is shaped by the distribution of star formation, and hence outflow, histories of the model galaxies, which reflect differences in the mass assembly histories of their halos.

Although the constant $Z_{\rm IGM}=10^{-3}\, Z_\odot$ model can match the average metallicity of the faint observed UFDs, such a model does not reproduce the observed scatter. Moreover, such a high metallicity floor of the accreted gas is excluded by the observation of UFD stars with metallicities further than one dex below this value \citep[e.g.,][]{Simon_2019}. Although a wide distribution with a central value near $\feh = -3$ would be consistent with observations, such a distribution does not seem to be produced by modern state-of-the-art galaxy formation simulations. 

Overall, we conclude that variability in the pre-enrichment of the IGM gas accreted by UFDs cannot explain the metallicity of these systems.
This conclusion is consistent with the results of modern simulations of dwarf galaxies in the cosmological context, which underpredict metallicities of UFDs by $\sim 0.5-1$ dex \citep[e.g., see Fig. 1 in][or Fig. 6 in \citealt{Go_etal_2025}]{Sanati_etal_2023}. 
This means that the gas accreted by UFD progenitors in these simulations is not enriched to the level that would result in a metallicity floor which would create a plateau at the average UFD stellar metallicity $\mathrm{[Fe/H]}\approx -2.5$. As we show next, this also indicates that the feedback-driven outflows in UFD progenitors in simulations are likely too strong. 

\subsection{Effect of mass loading factor on the stellar metallicity of UFDs}
\label{sec:outflow_modulation}

Given that the metallicity distribution of accreted gas in cosmological simulations does not explain the stellar metallicity of observed UFDs and, specifically, the plateau in the luminosity--metallicity relation at $\Mv\gtrsim -5$, this plateau is therefore likely to be driven by the metallicity regulation by internal enrichment and feedback-driven outflows \citep[e.g.,][]{Peeples.Shankar.2011,Muratov.etal.2017,Chisholm_etal_2018,GRUMPY}.  

The slope of the luminosity--metallicity relation in \texttt{GRUMPY} models is primarily determined by the adopted slope, $\eta_{\rm p}$, of the outflow mass loading factor, $\eta(M_*)$ (see eq.~\ref{eq:eta} above). The normalization at $M_\star\gtrsim 10^6\, M_\odot$ is set by $\eta_{\rm norm}$, while $\eta_{\rm max}$ sets an upper limit on $\eta$, flattening the relation. This in turn sets the value of the ``plateau'' in the stellar mass--metallicity relation at smaller values of $M_\star$.

The relatively tight bright-end luminosity--metallicity relation constrains most of the parameters of equation \ref{eq:eta}. As shown in Figure \ref{fig:mag-metal_mw-like}, the model reproduces the relation for galaxies of $\Mv\lesssim -6$ well using these parameters. The metallicity of UFD galaxies at lower luminosities 
is approximately constant on average and such average behavior can be matched if the average metallicity of the accreted gas was set to $Z_{\rm IGM}\approx 10^{-3}\, Z_\odot$ or greater. As was shown above, however, the accreted gas metallicity range predicted by several state-of-the-art simulations is consistent with average an order of magnitude lower.
In this case, the average metallicity of UFD galaxies can be matched by decreasing the \etamx\, value, which results in retention of a larger fraction of the metals which model UFDs produce within their ISM. 

In practice, we can expect that both star formation and feedback in individual galaxies exhibit stochastic fluctuations around the mean relation such as eq.~\ref{eq:eta}, especially as stellar mass decreases to $M_\star\lesssim 10^3$ where only a few supernovae are expected to occur throughout each galaxy's evolution. 
Thus, the outflows in individual galaxies (and their $\eta_{\rm max}$ values) should vary depending on the details of galaxy mass growth and star formation histories. Such variations would naturally lead to a scatter in the average metallicities of galaxies. In what follows, we do not attempt detailed modeling of such scatter given the relatively sparse observational constraints on its magnitude and functional form of the average metallicity distribution. Instead, we evaluate the range of $\eta_{\rm max}$ values that correspond to the observed range of UFD metallicities for  $\Mv\lesssim -6$, which gives us an idea of what outflow rates are required to match their metallicities. 

To examine the effect of outflows on the $\Mv-\rm [Fe/H]$ relation, we carry out the \texttt{GRUMPY} model calculation assuming constant metallicity of accreted gas $Z_{\rm IGM}=10^{-4}\, Z_\odot$ but with different values of the maximum mass loading factor, $\eta_{\rm max}$. As we showed above, the constant $Z_{\rm IGM}=10^{-4}\, Z_\odot$ model produced a median $\Mv-\rm [Fe/H]$ relation similar to those in which the accreted gas metallicity was sampled from the distributions measured in simulations, and is thus sufficient for this purpose. We also explicitly verified that the results below are similar for models where we sample accreted IGM metallicity from the distributions derived from simulations. 
As we saw above, for $Z_{\rm IGM}$ values that are so low and for the standard assumptions about the initial mass function of stars and their chemical yields, the metallicity of galaxies of $\Mv>-6$ cannot be explained by the metallicity of accreted gas. Therefore, exact values of $Z_{\rm IGM}$ are not particularly important.

\begin{figure}
    \centering
    \includegraphics[width=1.0\linewidth,trim= 16 16 16 16, clip]{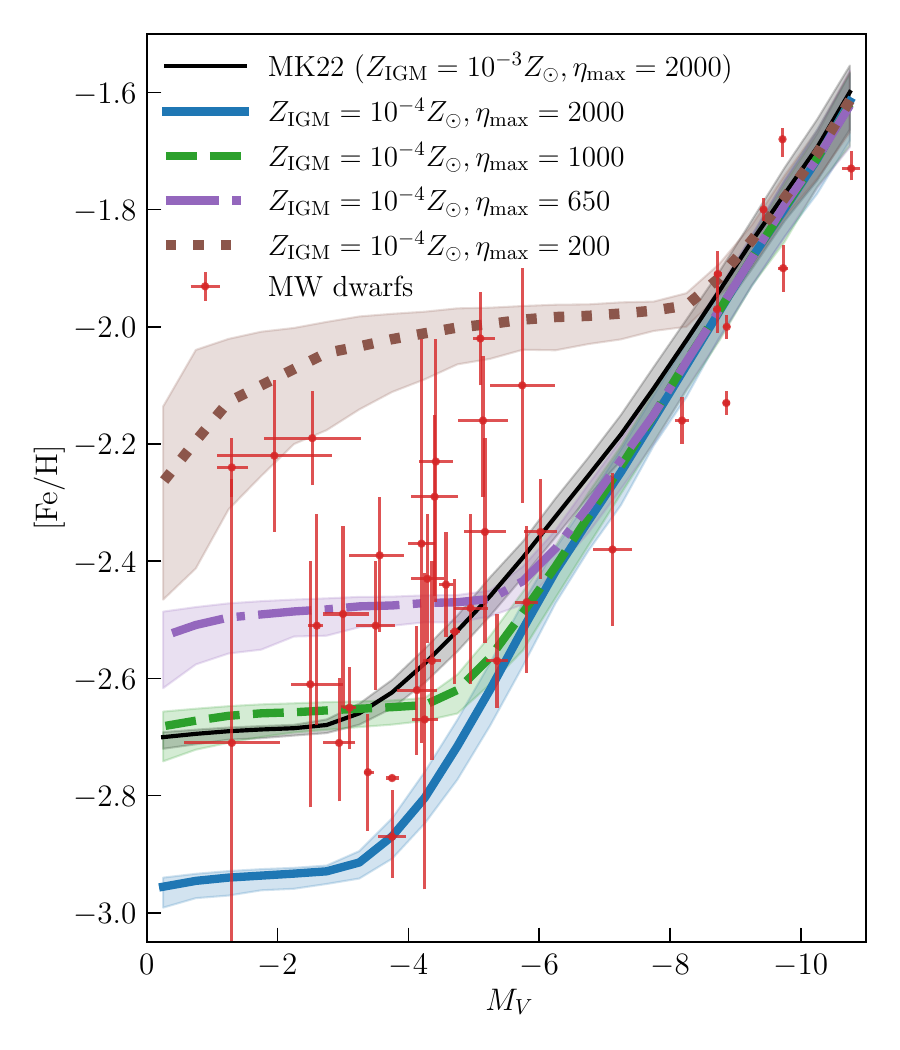}
    \caption{Same as Figure \ref{fig:mag-metal_mw-like}, however the \texttt{GRUMPY} model galaxies shown now vary {\etamx} with fixed $Z_{\rm IGM}=10^{-4}\, Z_\odot$, a value which produces outputs consistent with the \zigm\ distributions categorized in all cosmological simulation-informed models studied in previous sections. Varying \etamx\ has a profound effect on the location of the plateau, and $200<\eta_{\rm max}<2000$ is sufficient to bracket the entire observed range in average metallicity. Reducing \etamx\ from the fiducial value of $2000$ to $1000$ almost exactly recovers the same plateau position as the $Z_{\rm IGM}=10^{-3}\, Z_\odot$ model, and the plateau of the $\eta_{\rm max} = 650$ model is well aligned with the median observed UFD metallicity of $\feh=-2.48$.}
    \label{fig:mag-metal_etamx_range}
\end{figure}

Figure \ref{fig:mag-metal_etamx_range} shows the luminosity--metallicity relation for different \etamx\ values. The choice of \etamx\ controls the stellar mass at which metallicity deviates from the bright-end slope of the relation and thus the metallicity of the plateau in the relation at low luminosities. The figure shows that \etamx\ of $2000$ produces a mass-metallicity curve that brackets the low end of the scatter of observed UFDs, while the $\eta_{\rm max}=650$ model produces a plateau well aligned with the median of the metallicities of observed UFDs at $\feh=-2.48$, and $\eta_{\rm max}=200$ produces metallicities close to the highest metallicities of observed UFDs. Therefore, galaxy-to-galaxy variations of mass loading factor can easily explain the observed variation of UFD metallicities if mass loading factors vary by an order of magnitude in the range of $\eta\sim 200-2000$. 

These values of $\eta_{\rm max}$ are generally smaller than the extrapolation of average $\eta(M_\star)$ scaling adapted from \citet{Muratov_etal_2015} to UFD stellar masses, but are more comparable to the values of mass loading factor derived by \citet{Pandya_etal_2021} in FIRE-2 simulations for most of the UFD mass range: $\eta \simeq 2500, 890, 110$ for the stellar masses of $10^2, 10^3, 10^5\ M_\odot$, respectively. For fixed stellar mass, an order of magnitude range in $\eta$ also appears consistent with the scatter of $\eta$ for individual galaxies over their history reported by \citet{Pandya_etal_2021}. 
We note also that modeling of the stellar metallicity distribution of the UFD Eridanus II by \citet{Sandford.etal.2024} assuming a model with constant $\eta$ also indicated a value of $\eta\sim 200$ for that galaxy, consistent with the range of \etamx values we find here.

Because $\eta$ is a function of stellar mass, variation in \etamx\ most strongly effects low mass, low luminosity, galaxies and converges for brighter galaxies. This consequence mirrors the behavior of scatter in observed metallicity measurements, while as shown in Figure \ref{fig:mag-metal-scatter}, variation in accreted IGM metallicity over the evolution of a galaxy can effect scatter in all galaxies roughly independent of $\Mv$.

\subsection{Comparison of model and observed stellar metallicity distributions in UFDs}
\label{sec:MDFs}

\begin{figure*}
    \centering
    \includegraphics[width=\linewidth,trim= 0 40 0 0, clip]{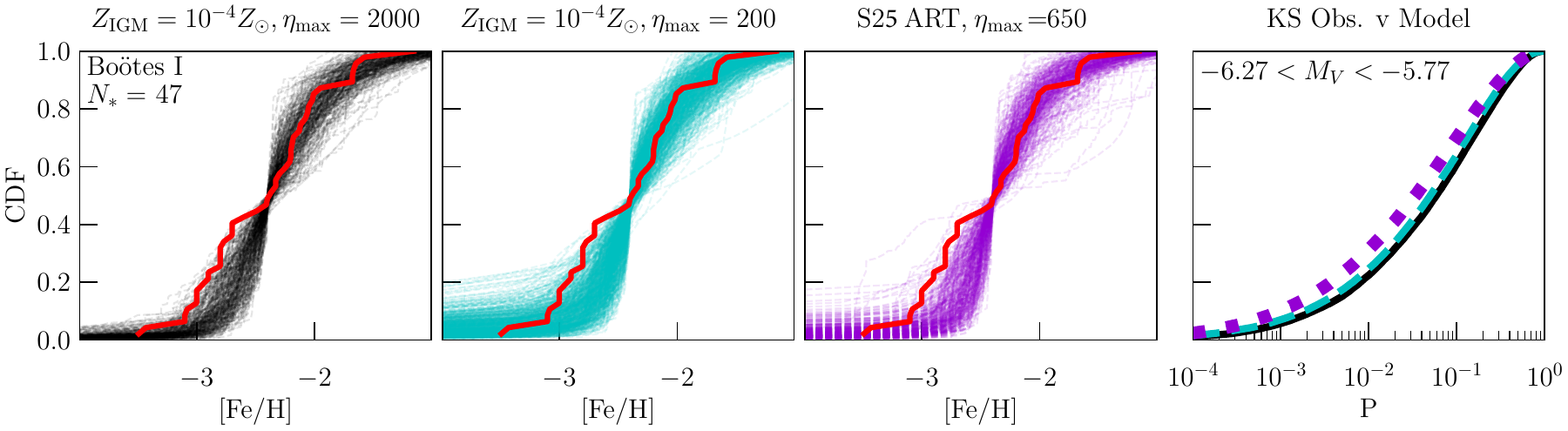}
    \includegraphics[width=\linewidth,trim= 0 40 0 21, clip]{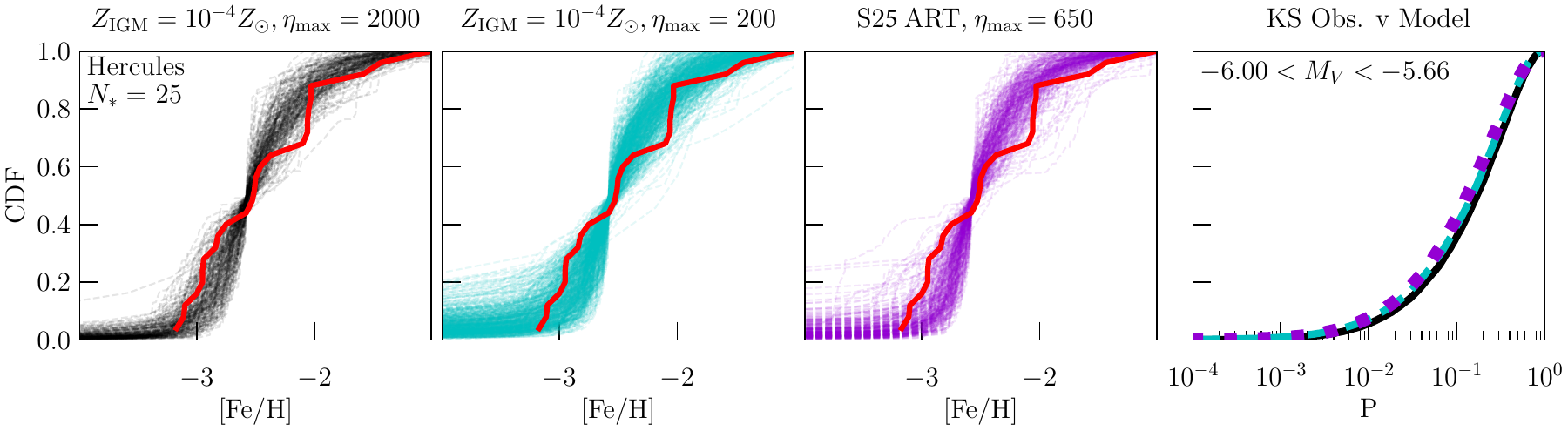}
    \includegraphics[width=\linewidth,trim= 0 40 0 21, clip]{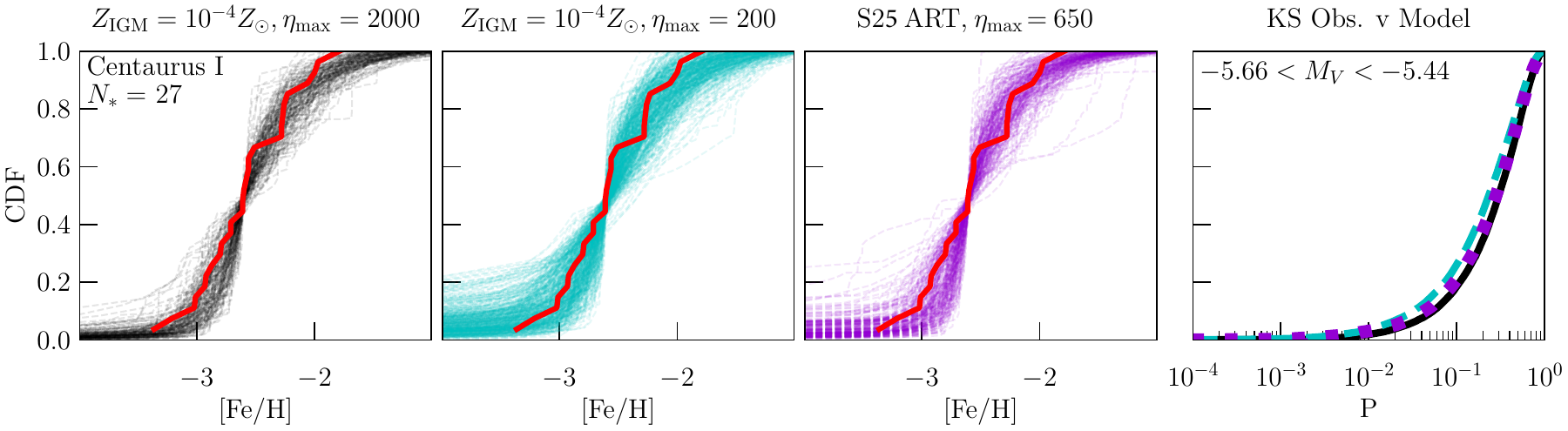}
    \includegraphics[width=\linewidth,trim= 0 40 0 21, clip]{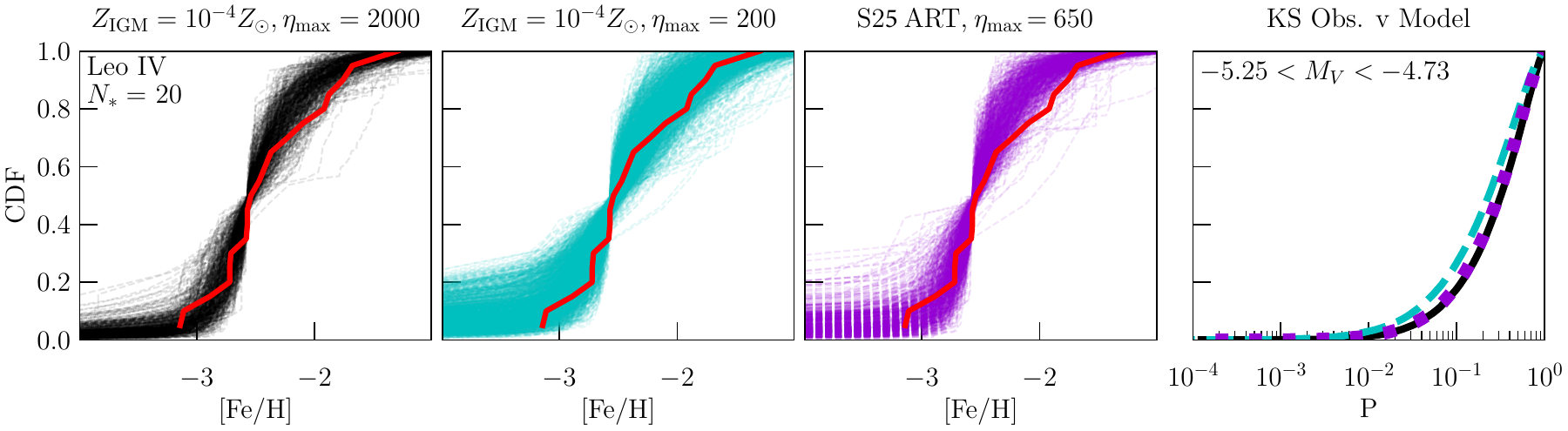}
    \includegraphics[width=\linewidth,trim= 0 0 0 21, clip]{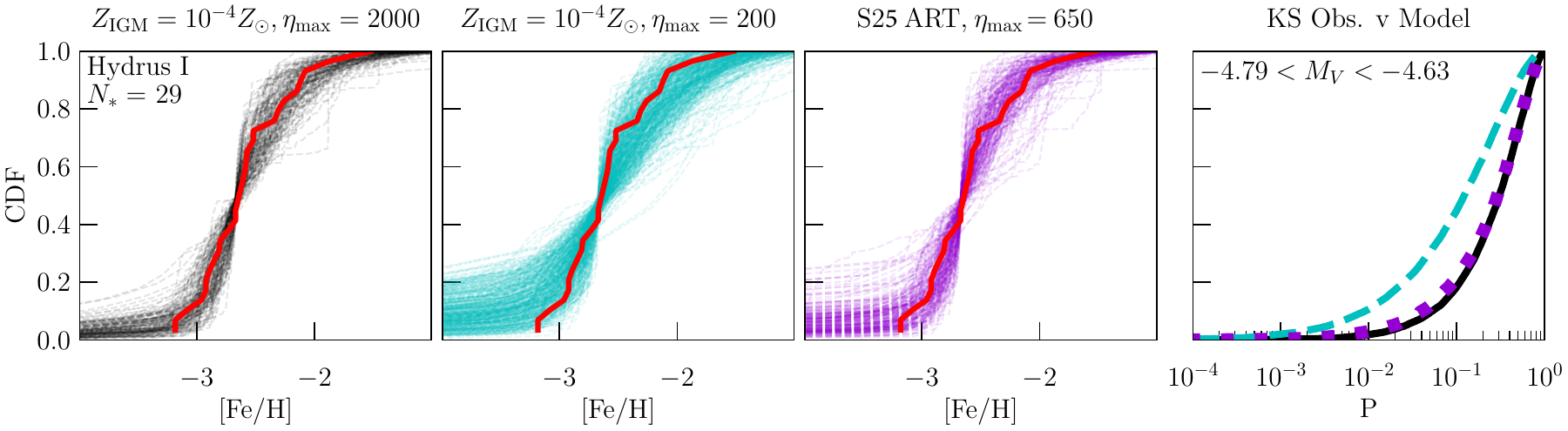}
    \caption{Comparing the shape of the MDF for models with various \zigm\ and \etamx\ prescriptions to observed UFD MDFs. Each dashed line in the left 3 columns is the MDF for an individual galaxy which falls within $1 \sigma$ of the magnitude of the observed galaxy to which it is compared. Here we compare only the shape of the MDF by offsetting the median value of the model MDF to match that of observation. Each stellar population of a model MDF is convolved with a stellar metallicity error from the observed galaxy in a bootstrap process. Each row shows a different observed UFD with MDF as a thick red line, while each column shows a different model. The number of spectroscopic observed stars making the MDF is given below the name of the UFD. The right column shows the cumulative distribution of Kolmogorov-Smirnov probability values comparing each model to the observed galaxy within a row. All model prescriptions produce MDF shapes that are nearly identical in comparison to observation, therefore no model is strongly preferred or disfavored. The \etamx$=200$ model has greater scatter and stronger tails than other distributions, as expected from the width of the contour in Figure \ref{fig:mag-metal_etamx_range}. Only in the faintest galaxies does this model stand out in KS probability, due to the weakly sampled tail of the observed MDFs in these galaxies.}
    \label{fig:mdf_comparison}
\end{figure*}

In addition to averaged metallicity properties for model galaxies, we examine how model results with different choices for \zigm\ and \etamx\ affect the cumulative distribution of stellar metallicities within individual galaxies. We compare model stellar metallicity distribution functions (MDFs) to the MDFs of observed UFD galaxies. Given that we compare the average stellar metallicities in the previous section, here we focus on comparing the shape of the distributions. We thus offset the model MDFs such that their median matches the median of the MDF of the observed galaxy to which it is compared.

To construct model MDFs, we track the evolution of single stellar populations (SSPs) within each galaxy using the Chempy chemical evolution model \citep{Chempy}. This model includes the time evolution of stellar populations and time-dependent enrichment and yields for individual elements. The model MDF is constructed as the cumulative distribution of \fehtxt\ of these SSPs, with the weight of contribution given by the stellar mass of each population at $z=0$ (corrected for mass loss).

To compare model MDFs to an observed UFD MDF, we select model galaxies with $\Mv$ within the $1 \sigma$ uncertainty of the $\Mv$ of an observed UFD. We perturb the metallicity of each SSP in each model galaxy with a random Gaussian number drawn from the pdf with standard deviation given by a randomly chosen observed stellar metallicity uncertainty from among stellar metallicity measurements of the observed comparison galaxy.

To quantify the relative success of different models in reproducing the observed MDF shape, we perform a Kolmogorov-Smirnov test comparing observed MDFs to the individual galaxy MDFs for each model. The resulting distribution of KS probability values for galaxies within a model gives a quantitative comparison of which model better reproduces the observed MDF shape. We compare models to $5$ observed UFD MDFs, which have metallicity measurements of at least $20$ stars and do not have significant overlap in $\Mv$. Figure \ref{fig:mdf_comparison} shows MDFs from models with $\Zigm=10^4\, Z_\odot$, $\eta_{\rm max} = 200\ {\rm and}\ 2000$, and a model with \zigm\ drawn from the S25-based \zigm\ distribution with $\eta_{\rm max}=650$. The first two span the range in \etamx\ required to bracket the observed luminosity--metallicity relation. The latter gives an intermediate value of \etamx\ which creates a luminosity--metallicity curve approximating the mean metallicity of UFDs and shows a non constant, simulation-informed \zigm.

The left 3 columns of Figure \ref{fig:mdf_comparison} show multiple bootstrap realizations of individual model galaxy MDFs compared to the observed MDF for a single UFD on each row.  
The figure shows that all models match the observed metallicity distributions reasonably well. This is because the shape of the observed metallicity distribution is currently dominated by metallicity uncertainties and differences between predicted model MDFs are smaller than the effect of these uncertainties. 

Comparing to the \zigm$=-4$, $\eta_{\rm max} = 2000$ model, it can be seen that both other models shown have a greater variety of individual MDF shapes. The $\eta_{\rm max} = 200$ model produces more MDFs with significantly stronger tails, especially at the lowest metallicities. This is consistent with the wider $1 \sigma$ contour in luminosity--metallicity for this model shown in Figure \ref{fig:mag-metal_etamx_range}.
The model with the S25-based \zigm\ distribution, which produces the strongest tail of high metallicities in galaxies of a given $\Mv$ (see Figure \ref{fig:mag-metal-scatter}) does not produce notably more MDFs skewed to high metallicity than other models.
Overall, there is no clear qualitative difference in the average MDF between models, or their success in matching the observed MDFs.

Quantitatively, the right column of Figure \ref{fig:mdf_comparison} shows that for all but the faintest of UFDs, the distribution of probabilities for the model and observational MDFs to be drawn from the same underlying distribution is similar among models, and the typical probability values are high. In the faintest galaxies shown, the $\eta_{\rm max} = 200$ model extends to lower overall KS probabilities than the other models. The fainter observed UFDs tend to have a weaker low-metallicity MDF tail than brighter UFDs, and especially weak compared to the prominent low-metallicity tail of the $\eta_{\rm max} = 200$ model. Nevertheless, the small number of stellar metallicity samples for the observed UFD does not fully sample this tail to probe this difference. Larger samples of stars with metallicity measurements and smaller metallicity uncertainties are required to probe differences imposed by different outflow rates in the full range of the models required by the observed scatter of average metallicities of UFDs.

\section{Discussion}
\label{sec:discussion}

\subsection{The role of Pop III stars in chemical enrichment of UFD stars}

Much of the current discussion of the pre-enrichment of the IGM in the literature is centered on the physics of very low metallicity and Pop III stars. In our study, the FIRE-2 and S25 simulations do not model Pop III stars and their contribution to feedback and enrichment, while the MEGATRON models do include explicit modeling of Pop III stars and their contribution to chemical enrichment. We find that there is very little difference in the luminosity--metallicity relation of dwarf galaxies when accretable IGM metallicity distributions from these simulations are used in the \texttt{GRUMPY} galaxy formation model. Likewise, there is no significant difference in the predicted metallicity distributions of stars resulting from these distributions. The exact chemical abundance of stars may differ due to Pop III yields, however we do not model the accretion of individual elements in detail.

At redshifts $z\gtrsim 8$, where UFDs are most likely to form the bulk of their stars, the distributions of IGM available to UFDs are qualitatively similar, in that very little gas is enriched above $\feh=-4$. The contribution of Pop III stars explicitly modeled in the MEGATRON simulations thus does not lead to substantial amounts of more enriched gas available compared to the FIRE-2 and S25 simulations. This is consistent with observations of the lowest metallicity damped Lyman $\alpha$ system at $z=3.08$, which has an upper limit on metallicity of $\mathrm{[Fe/H]}\lesssim -3.66$ and constraints on the abundances of C, Al, and Si consistent with the origin in a single Pop III supernova \citep[][see also \citealt{Erni_etal_2006,Cooke_etal_2011}]{Welsh_etal_2023}. 

It is beyond the scope of this analysis to identify the contribution of Pop III stars to IGM enrichment in the MEGATRON simulations, but we note that the small peak in \zigm\ near $\feh=-6$ in the metallicity distribution extracted from these runs (see Figure \ref{fig:zigm_cdf}) is likely a signature of the Pop III enrichment because that metallicity is assumed to correspond to the transition between Pop II and Pop III populations in the MEGATRON model at $10^{-6}\, Z_\odot$ \citep{Katz_etal_2024_MEGATRON}. Our results, however, indicate that the final metallicity distribution of stars in model UFDs is insensitive to the distribution of gas with $\Zigm<10^{-4}\, Z_\odot$. Therefore any contribution of Pop III stars to $\Zigm$ at such low metallicities does not affect our results. 

Although the effect of the Pop III stars on the metallicity of the IGM accretable by UFDs may be limited, self-enrichment by Pop III stars in the UFD progenitors may potentially be more important \citep[e.g.,][]{Salvadori_Ferrara_2009,Frebel_Bromm_2012,Frebel_Norris_2015}. Indeed, 
\citet{Wise_etal_2012} showed that Pop III stars can potentially self-enrich gas within halos of mass $\gtrsim 3\times 10^7\, M_\odot$ at $z\gtrsim 7$ expected to be progenitors of $z=0$ UFDs to $Z\sim 10^{-3}\, Z_\odot$. Thus, the Pop III contribution to the self-enrichment of 
UFD progenitors, not included in the \texttt{GRUMPY} model, can potentially affect the low-metallicity tail of the stellar metallicity distribution of UFDs and the average metallicity of the galaxies with lowest metallicities. We note, however, that most UFDs have average metallicity a factor of $\approx 3-10$ larger than $10^{-3}\, Z_\odot$. The level of enrichment predicted in the simulations of \citet{Wise_etal_2012} therefore cannot explain these metallicities or the lack of a trend in the luminosity--metallicity relation at $\Mv>-6$.

Likewise, \citet{Sanati_etal_2023} examine the effect of Pop III self-enrichment on the metallicity of UFD dwarfs in their simulations using different assumptions about their IMF and nucleosynthetic yields. They find that although the inclusion of Pop III stars does raise $\mathrm{[Fe/H]}$ in the simulated galaxies by $\approx 0.2-0.5$ dex depending on whether enrichment by pair instability supernovae (PISN) is included, even the simulations with largest enrichment produce average UFD metallicities that are more than $0.5$ dex below metallicities of observed low luminosity UFDs. Their simulations also do not produce the large scatter in average metallicity observed at $\Mv<-6$. \citet{Mead_etal_2025} find that varying the Pop III IMF, without explicit inclusion of PISN has little impact on the mass-metallicity relation at the redshifts simulated ($z>14$), suggesting that increased metal yields from high mass stars are balanced by lower efficiency in enriching the ISM.

Conversely, \citet{Prgomet_etal_2022} find that introducing a metallicity dependent IMF to a simulation of a single dwarf galaxy from the EDGE simulations is capable of producing UFDs with stellar metallicities $\feh>-2.5$ in agreement with the observed luminosity--metallicity relation. A greater number of high mass stars increases burstiness, regulating star formation and reducing stellar mass for a given halo mass compared to their fiducial model. More high mass stars also increase iron yields commensurately to produce a relatively flat trend for permutations of this galaxy. A similar result was found with the inclusion of radiative feedback in addition to supernovae.

Our results and the results of the studies discussed above strongly indicate that neither enrichment of the IGM by other galaxies in the environment of UFD progenitors nor their self-enrichment by Pop III stars are sufficient to explain the average metallicities of observed UFDs and their scatter.

\subsection{The mass loading factor}

Our results indicate that metallicity measurements in UFDs can constrain outflow properties in the progenitors of these galaxies \citep[see also][]{GRUMPY,Sandford.etal.2024}. In particular, the observed average metallicities indicate outflow mass loading factors in the range of $\eta\approx 200-2000$. 
In cosmological simulations, relations for mass loading factor predict values of several hundred to several thousand when extrapolated to the low stellar mass range of UFDs \citep{Muratov_etal_2015,Pandya_etal_2021,Mitchell_etal_2020,Nelson_etal_2019} -- broadly consistent with the deduced range up to the lowest mass UFDs. We note that a similar range of $\eta$ is required to reproduce the existing constraints on the stellar mass--halo mass relation of dwarf galaxies \citep[e.g.,][]{GRUMPY,MK22}.

Each of \citep{Muratov_etal_2015,Pandya_etal_2021,Mitchell_etal_2020,Nelson_etal_2019} adopt a slightly different definition for which gas is considered outflowing, incorporating different cuts on radial velocity and distance as well as inclusion of various other parameters. \citet{Nelson_etal_2019} demonstrate that the stellar mass scaling of $\eta$ can be sensitive to the definition of outflow in simulations.

However, attempts to estimate mass loading factors in ongoing outflows of nearby observed dwarf galaxies tend to derive significantly lower values from $\eta\sim 0.2-20$ \citep{Heckman_etal_2015,Chisholm_etal_2017,McQuinn_etal_2019,Kado_Fong_etal_2024} to $\eta\sim 0.02$ \citep{Marasco_etal_2023}. Comparatively low values of $\eta$ are also estimated for observed galaxies at higher $z$ \citep{Concas_etal_2022,Carniani_etal_2024}.
These estimates, however, are for galaxies of much larger stellar mass ($M_\star \gtrsim 10^7\, M_\odot$) than UFDs and also generally show indications that $\eta$ increases with decreasing stellar mass similar to theoretical predictions \citep[see, however,][]{McQuinn_etal_2019}. One caveat is that such estimates rely on indirect tracers of outflowing mass, which may underestimate the outflow rate if most of the mass is in a phase very different from that probed by observations. The estimates of the mass loading factors based on the modeling of metallicity distribution and chemical abundances in UFDs, on the other hand, deduce values of $\eta$ consistent with the values estimated here \citep{Johnson_etal_2023,Alexander_etal_2023,Sandford.etal.2024}

Redshift evolution of $\eta$ is negligible in the FIRE-1 and FIRE-2 simulations for $0 < z < 4$ \citet{Muratov_etal_2015, Pandya_etal_2021}, while \citet{Mitchell_etal_2020} find a weak redshift dependence at $z<3$, but no evolution at $z>3$. Therefore, $\eta$ is not expected to evolve significantly with redshift, and low redshift measurements can be compared to our results, which most strongly impact UFDs at $z \gtrsim 8$ when they are expected to be star forming.

The expected scatter in $\eta$ is not well understood. Theoretical analyses that focused on the long-time average $\eta$ measurements find a rather small scatter of $\eta$ at a given stellar mass \citep[e.g.,][]{Muratov_etal_2015,Pandya_etal_2021}. In contrast, \citet{Pandya_etal_2021} also report measurements of $\eta$ for individual starburst events within a galaxy, which appear to agree with the order of magnitude variation in the maximum value \etamx\ we find is needed to reproduce the range of observed UFD metallicity scatter. This may imply that the mechanisms that cause variation in \etamx\ already operate in higher stellar mass dwarfs, but the scatter of individual starbursts averages out in massive galaxies experiencing multiple bursts. UFDs, on the other hand, may form their stars in a small number of bursts and burst-to-burst variations of $\eta$ may translate to a large variation of average metallicity in these systems. Additional scatter may arise in the smallest UFDs due to stochasticity of feedback due to a small number of massive stars at any given time. 

We find that on average, $\eta\approx\mathrm{const}$ is required to produce flattening in the luminosity--metallicity relation.
\citet{Rey_etal_2025}, for example, find that the inclusion of radiative feedback in addition to SN feedback in their EDGE simulations reduces $\eta$ and leads to higher metallicity in UFDs, making a better fit to luminosity--metallicity observations. In simulations with radiative feedback, the time-averaged $\eta$ is relatively constant in stellar mass for their sample of UFDs, with the $68\%$ interval between a few tenths to several hundred over the cosmological history of a UFD. This agrees with $\eta\approx\rm const$ as an explanation for the flattening of the $\mathrm{[Fe/H]-M_V}$ relation in the UFD regime, and broadly with our required range of \etamx\ values.

Rather than regulating star formation with different modes of ISM heating, the outflow wind model in TNG50 determines wind injection velocity by the local dark matter velocity dispersion, subject to a minimum velocity requirement, therefore the winds launched at the injection scale reach a well defined maximum that leads to flattening in $\eta$ at this scale \citep{Pillepich_etal_2018_TNG,Nelson_etal_2019}. However, as \citet{Nelson_etal_2019} show, the emergent outflows at larger radii do not necessarily show the same flattening, subject to the radial velocity and distance used to define galactic outflow.

We note that our compilation of observed UFD mean metallicities are taken from spectroscopically derived values in the Local Volume Database \citep[][]{Pace_lvdb}. These are compiled from a range of studies, but predominantly originate from medium-resolution spectroscopic observations where metallicities are derived from the near-infrared calcium triplet lines \citep[e.g.][]{Li_etal_2017,Longeard_etal_2018,Jenkins_etal_2021_MDF,Bruce_etal_2023}. Metallicity calibrations from these features are reliable down to $\feh = -4.0$ with a minimum precision of $\sim 0.16$ dex \citep[e.g.][]{Carrera_etal_2013} that varies based on the S/N of the observed spectrum of each star. This metallicity calibration, and other typical spectroscopic calibrations (e.g., the Ca II K line; \citet{Beers_etal_1999}) or methods (i.e., high-resolution spectroscopy) do not impose an artificial metallicity floor at $\feh \sim -3.0$. Accordingly, it is unlikely that the leveling of the metallicity-luminosity relation is due to an artifact in the derived metallicities.

Our results and those of studies discussed here highlight the importance of feedback-driven outflows in determining the metallicity of UFDs, but further work is required to understand the mechanisms which create the scatter in $\eta$ which we find required to produce the large range in observed UFD metallicities.

\subsection{The mass loading factor and the $M_\star-M_{\rm h}$ relation}

\begin{figure}
    \centering
    \includegraphics[width=1.0\linewidth]{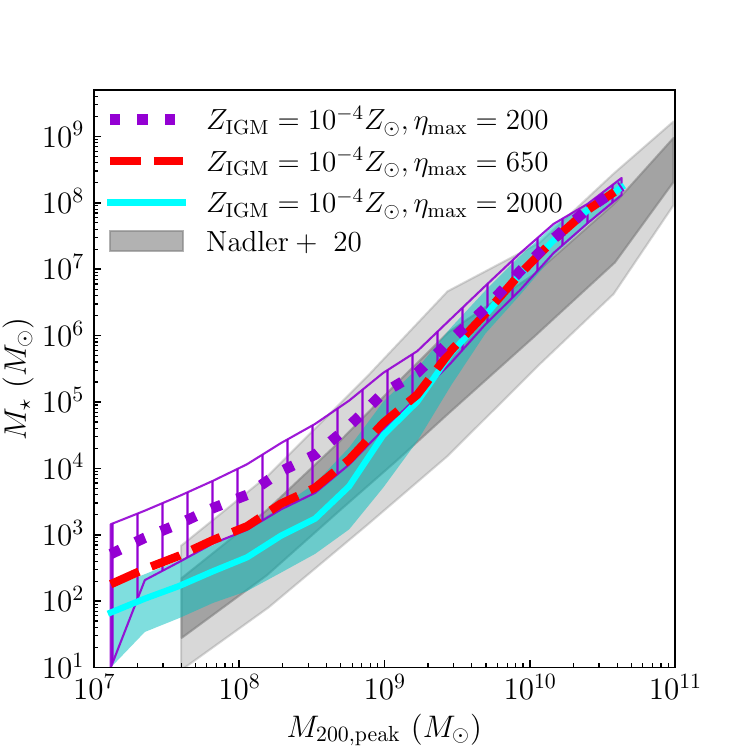}
    \caption{Stellar to halo mass relation for \texttt{GRUMPY} model galaxies with \etamx\ varied over the range which brackets the spread of observed values in the luminosity--metallicity relation, with \zigm\ consistent with cosmological simulations. The grey shaded region represent 1 and 2 $\sigma$ equivalent constraints of the relation from \citet{Nadler_etal_2020}, while the colored shaded or hatched regions correspond to $68\%$ regions of the \texttt{GRUMPY} models shown. To reduce clutter, the contour for $\eta_{\rm max}=650$ is not included, but it is comparable to $\eta_{\rm max}=2000$. Models with values of $\eta_{\rm max}=650$ and $2000$ are consistent with constraints over nearly their entire range, and the $\eta_{\rm max}=200$ model does not drastically deviate until within the UFD mass regime regime. In reality, we expect \etamx\ to drawn from a distribution covering this range, so $\eta_{\rm max}=200$ applied to every galaxy in the model represents the most extreme upper bound. Such a distribution peaked near $\eta_{\rm max}=650$ appears generally consistent with constraints of \citet{Nadler_etal_2020}.}
    \label{fig:stellar-halo-mass}
\end{figure}

The fiducial outflow parameters and parameterization of $\eta$ discussed in Section \ref{subsec:model_grumpy} have been shown to reproduce several observed properties and scaling relations of dwarf galaxies \citep{GRUMPY,MK22}. In this work, we have only varied the maximum mass loading factor $\eta_{\rm max}$ to determine the range of its values required to match the observed luminosity--metallicity relation in the UFD regime. In the model with $\eta_{\rm max}=200$, the characteristic values of $\eta$ are reduced by a factor of several for most of the model UFDs compared to the fiducial values. 
This increases both the metallicity and stellar mass of model galaxies. One should thus consider whether the stellar mass increase is consistent with existing constraints on the $M_\star-M_{\rm h}$ relation for UFD galaxies.

Figure \ref{fig:stellar-halo-mass} shows the stellar mass -- peak halo mass (the largest halo mass during evolution to $z=0$ of a host halo) relation for models of $\Zigm=10^{-4}\, Z_\odot$, $\eta_{\rm max} = 200,650,2000$. We compare the model relation to constraints for observed UFD population around Milky Way \citep{Nadler_etal_2020}. The $\eta_{\rm max} = 2000$ model corresponds to the fiducial model of \citet{MK22}, and is therefore consistent with constraints for the entire UFD mass range. The relation in the $\eta_{\rm max} = 650$ model is close to the fiducial curve and is also in good agreement with observational constraints. The $\eta_{\rm max} = 200$ model relation is $1\sigma$ higher than the median of the constraint at $M_\star \lesssim 10^4\ M_\odot$ and is $2\sigma$ higher than the median constraint at $M_\star \lesssim 10^3\ M_\odot$. Therefore, this model may overpredict the stellar masses of the faintest UFDs and the abundance of UFD satellites around the Milky Way. 

Note, however, that $\eta_{\rm max}=200$ is the smallest value required to bracket the scatter of metallicities of observed UFDs. Most UFDs will have larger values of $\eta$ and correspondingly smaller stellar masses with the average value of $\eta_{\rm max}\approx 650$.
Figure~\ref{fig:stellar-halo-mass} shows that such a model is in reasonable agreement with the constraint of \citet{Nadler_etal_2020}.

Finally, we note that in the semi-analytic models of the kind used in this study, the mass loading factor $\eta$ is assumed to correspond to the star formation rate averaged over some suitably long time scale. Star formation in dwarf galaxies is generally expected to be bursty due to large scatter in the relation between star formation rate, gas mass and stellar mass \citep[e.g.,][]{Pan_Kravtsov_2023}, which may lead to large fluctuations of $\eta$ and corresponding scatter in metallicity and stellar mass. Significant metallicity scatter in the UFD regime may thus reflect the bursty star formation in these galaxies.

Interestingly, modern simulations of galaxy formation generally underpredict both the average metallicity and the metallicity scatter of UFDs \citep[e.g., see Fig. 1 in][]{Sanati_etal_2023}. This may indicate the inapplicability of standard recipes for star formation and feedback in the UFD regime, underscoring the role of the faintest galaxies as a testing ground for such models.

\section{Summary and Conclusions} 
\label{sec:conclusion}

We used a combination of the IGM metallicity distributions derived from several state-of-the-art cosmological simulations and a semi-analytical model of UFD galaxy formation to explore a range of models with different assumptions about the accreted IGM metallicity and outflow mass loading factors. We compare the luminosity--metallicity relation of the model galaxies and their stellar metallicity distribution functions (MDFs) to existing measurements for observed UFDs to gauge the relative importance of the IGM enrichment and metal regulation by outflows in setting UFD metallicities.
Our results and conclusions are summarized as follows. 

\begin{itemize}
\item[1.] We examine the metallicity distribution of the IGM gas that would be accretable by UFD progenitor galaxies at $5<z<10$ in several state-of-the-art simulations of galaxy formation with different treatments of the gas thermodynamics and radiative transfer, star formation, Pop III modelling, and stellar feedback. We find that although the distributions are fairly broad, most of the available IGM gas has low metallicity with only a few per cent of the IGM available to UFDs enriched to $\feh \ge -4$ (see Figure \ref{fig:zigm_cdf}). 

\item[2.] We show that when the IGM metallicity distributions measured in simulations are sampled to set the metallicity of the gas accreted by UFD progenitors in the semi-analytic model of galaxy formation, the average stellar metallicities of model galaxies are significantly below the metallicities of observed UFDs. The scatter of average metallicities in the model is also significantly smaller than scatter exhibited by observed galaxies (see Figure \ref{fig:mag-metal_mw-like}). This result does not 
depend on the details of the IGM metallicity distribution at $\feh \lesssim -4$.

\item[3.] In agreement with previous studies, we find that inclusion of Pop III modeling does not change the IGM metallicity distribution of the IGM above $\mathrm{[Fe/H]\approx -4}$. The metallicity increase of UFDs due to Pop III enrichment of the environment therefore cannot account for the average metallicity of these systems or the full range of metallicity values observed.

\item[4.] Our results indicate that the IGM enrichment is not the dominant factor in shaping the scatter seen in the average stellar metallicities of UFDs.
The metallicity of UFD stars is determined primarily by the internal interplay between enrichment by its stars and loss of metals in feedback-driven outflows. 

\item[5.] We examine the results of models with different values of the maximum outflow mass loading factor $\eta_{\rm max}$ and show that the full range of average stellar metallicities of UFDs at $M_V<-7$ can be reproduced if the maximum mass loading factor varies in the range $200\lesssim\eta_{\rm max} \lesssim 2000$, with $\eta_{\rm max} \sim 650$ corresponding to the median metallicity of observed UFDs ($\feh\approx-2.5$) (see Figure \ref{fig:mag-metal_etamx_range}). We show that model stellar mass--halo mass relation for this range of $\eta_{\rm max}$ values is also consistent with existing constraints (see Figure \ref{fig:stellar-halo-mass}).

\item[6.] We also compare stellar metallicity distributions within  individual observed UFDs and model galaxies with different assumptions about \zigm\ and \etamx\ values. We find that all considered models are in reasonable agreement with observed metallicity distributions (see Figure \ref{fig:mdf_comparison}). This is because current uncertainties of metallicity measurements are larger than predicted variations of model metallicity distributions within the range of models we considered. 

\end{itemize}

The range of  \etamx\ values deduced in our study is generally consistent with extrapolations of scaling relations of mass loading factor with stellar mass, $\eta(M_\star)$ measured in cosmological simulations  \citep{Muratov_etal_2015,Angles_Alcazar_etal_2017,Pandya_etal_2021,Mitchell_etal_2020}. Our conclusions, however, imply that the ``flattening'' of the observed luminosity--metallicity relation and increase of metallicity scatter in the UFD regime signal the breakdown of tight $\eta(M_\star)$ scaling in these systems. 
 
Our results therefore, highlight the need for further theoretical exploration of the star formation--feedback cycle and outflow properties in the smallest UFD galaxies to understand the processes responsible for the variation in mass loading factor required to reproduce the average metallicity and its scatter for observed UFDs.

\section*{Acknowledgments}
We thank Alexander P. Ji and the Caterpillar collaboration for providing halo evolution tracks of the Caterpillar simulations used in this study. VW and AK were supported by the NASA ATP grant 80NSSC20K0512 and the National Science Foundation grant AST-2408267. AC is supported by a Brinson Prize Fellowship at KICP/UChicago. Support for VS was provided by Harvard University through the Institute for Theory and Computation Fellowship. We have used the Astrophysics Data Service (\href{http://adsabs.harvard.edu/abstract_service.html}{\tt ADS}) and \href{https://arxiv.org}{\tt arXiv} preprint repository extensively during this project and the writing of the paper.

\section*{software}
Analyses presented in this paper were greatly aided by the following free software packages: {\tt NumPy} \citep{NumPy2015,numpy}, {\tt SciPy} \citep{SciPy}, {\tt Matplotlib} \citep{matplotlib}, {\tt FSPS} \citep{fsps} and its Python bindings package {\tt Python-FSPS}\footnote{\href{https://github.com/dfm/python-fsps}{\tt https://github.com/dfm/python-fsps}}. 

\section*{Data Availability}
We used FIRE-2 simulation public data  \citep[][]{Wetzel_etal_2023_FIRE_pubdata}, which are part of the Feedback In Realistic Environments (FIRE) project, generated using the Gizmo code \citep{Hopkins_2015_GIZMO_code} and the FIRE-2 physics model \citep{Hopkins_etal_18_FIRE_code}.
Halo catalogs from the Caterpillar simulations are available at \href{https://www.caterpillarproject.org/}{\tt https://www.caterpillarproject.org/}. The \texttt{GRUMPY} model pipeline is available at \url{https://github.com/kibokov/GRUMPY}. The data used in the plots within this article are available on request to the author.

\bibliographystyle{mnras}
\bibliography{manuscript}

\end{document}